\documentstyle[12pt]{article}
\newlength{\mytopmargin}
\newlength{\myleftmargin}
\def\zz{\rlx\hbox{\small \sf Z\kern-.4em Z}}

\newcommand{\ml}{\langle}
\newcommand{\mg}{\rangle}

\begin{document}

\vspace{1cm}
\noindent
\begin{center}{   \large \bf
Analytic properties of the structure function for the \\
one-dimensional one-component log-gas
}  
\end{center}
\vspace{5mm}

\noindent
\begin{center} 
 P.J.~Forrester$^*$, B.~Jancovici$^\#$ and D.S.~McAnally$^*$\\

\it $^*$Department of Mathematics
and Statistics, University of Melbourne,\\ Parkville, Victoria
3052, Australia\\[.1cm]
$^\#$Laboratoire de Physique Th\'eorique, Universit\'e Paris-Sud,\\
91405 Orsay Cedex, France\footnote{Laboratoire associ\'e au Centre
National de la Recherche Scientifique - URA D0063}
\end{center}

\begin{center}
Dedicated to R.J.~Baxter on the occasion of his 60$^{\rm th}$ birthday.
\end{center}
\vspace{.5cm}

\small
\begin{quote}
The structure function $S(k;\beta)$ for the 
one-dimensional one-component log-gas is the Fourier
transform of the charge-charge, or equivalently the density-density,
correlation function. We show that for $|k| < {\rm min} \,(2\pi \rho,
2 \pi \rho \beta)$, $S(k;\beta)$ is simply related to an analytic
function $f(k;\beta)$ and this function satisfies the functional equation
$f(k;\beta) = f(-2k/\beta;4/\beta)$. It is conjectured that the coefficient
of $k^j$ in the power series expansion of $f(k;\beta)$ about $k=0$ is of the
form of a polynomial in $\beta/2$ of degree $j$ divided by $(\beta/2)^j$.
The bulk of the paper is concerned with calculating these polynomials 
explicitly up to and including those of degree 9. It is remarked that the
small $k$ expansion of $S(k;\beta)$ for the two-dimensional one-component
plasma shares some properties in common with those of the 
one-dimensional one-component log-gas, but these
break down at order $k^8$.
\end{quote}

\vspace{.5cm}

\section{Introduction}
\setcounter{equation}{0}
The one-component log-gas, consisting of $N$ unit charges on a circle of
circumference length $L$ interacting via the two-dimensional Coulomb
potential $\Phi(\vec{r},\vec{r}\,') = - \log |\vec{r} - \vec{r}\,'|$,
is specified by the Boltzmann
factor
\begin{equation}\label{1}
A_{N,\beta} \prod_{1 \le j < k \le N}
| e^{2 \pi i x_k / L} - e^{2 \pi i x_j / L}|^\beta, \qquad 0 \le x_j \le L.
\end{equation}
The constant $A_{N,\beta}$, which plays no role in the calculation of
distribution functions, results from scaling the radius of the circle out
of the logarithmic potential, and also includes the particle-background
and background-background interactions (a uniform neutralizing background
is imposed for thermodynamic stability).
The thermodynamic limit
$N,L \to \infty$,$N/L = \rho$ (fixed) is taken, which gives an infinite
system on a straight line with particle density $\rho$.
This system was first studied because of its relation to
the theory of random matrices \cite{Me91}. The thermodynamic functions
were obtained. The pressure $P$ has the simple form
\begin{equation}\label{pressure}
\beta P=[1-(\beta /2)]\rho
\end{equation}
at any inverse temperature $\beta$. However, exact (simple) forms for the
correlation functions were obtained by the pioneers only for the special
temperatures corresponding to $\beta=1,2,4$ (See Section 5). More
recently, exact  expressions for the two-body density were derived for
arbitrary even integer $\beta$ \cite{Fo93aa} and then for arbitrary
rational $\beta$ \cite{Ha95}. Unfortunately, these latter exact
expressions are complicated multivariable integral representations which
cannot be easily used as such for actual computations. The purpose of
the present paper is to obtain explicit small $k$ expansions for the
structure function (the Fourier transform of the two-body density).

The log-gas is an example of a system interacting via the $d$-dimensional
Coulomb system (here $d=2$) but confined to a domain of dimension $d-1$.
It therefore exhibits universal features --- that is features independent of
microscopic details such as any short range potential between charges 
or the number of
charge species --- characteristic of Coulomb systems in this setting
\cite{Ja82}. One
universal feature is the existence of an algebraic tail in the leading
non-oscillatory term of the large-distance asymptotic expansion of the
charge-charge correlation function. For general charged systems in their
conductive phase, interacting via the two-dimensional Coulomb potential
in a one-dimensional domain, this is predicted to have the form
\cite{FJS83}
\begin{equation}\label{2}
- {1 \over \beta (\pi r)^2},
\end{equation}
where $r$ is the distance. For the
one-component log-gas, (\ref{2}) can be verified for all $\beta$
rational \cite{FZ96}.

The verification is possible because the charge-charge correlation function
(which for a one-component system is the same as the density-density
correlation)
is known explicitly for $\beta$ rational \cite{Ha95} (see (\ref{Ha})
below). In this work we further analyze the properties of the
structure factor $S(k;\beta)$ (Fourier transform of the charge-charge
correlation) for the one-component log-gas. In particular we are interested
in the $\beta$ dependence of the coefficients in the small $k$
expansion of $S(k;\beta)$.

The large distance behaviour (\ref{2}) is equivalent to the small $k$
behaviour
\begin{equation}\label{3}
S(k;\beta) \sim {|k| \over \pi \beta}.
\end{equation}
Furthermore, by making use of the equivalence of the charge-charge and
density-density correlation in the one-component log-gas,
together with the exact equation of state
the second order term in (\ref{3}) has been predicted for general $\beta$
\cite{FJ97}, giving
\begin{equation}\label{4}
S(k;\beta) \sim {|k| \over \pi \beta} 
+ {(\beta / 2 - 1) k^2 \over (\pi \beta)^2 \rho } +
O(|k|^3).
\end{equation}

Let 
\begin{equation}\label{4+}
f(k;\beta) := {\pi \beta \over |k|} S(k;\beta), \quad 0 < k < {\rm min}\,
(2 \pi\rho, \pi \beta\rho)
\end{equation}
and define $f$ for $k < 0 $ by analytic continuation
(we will see below that $f(k;\beta)$ is analytic for
$0 \le |k| < {\rm min}\, (2 \pi \rho, \pi \beta \rho)$).
In Section 2 we use the exact result (\ref{Ha}) below to derive the
functional equation
\begin{equation}\label{5}
f(k;\beta) = f\Big (-{2 k \over \beta}; {4 \over \beta} \Big )
\end{equation}
The
simplest structure consistent with (\ref{5}) is
\begin{equation}\label{6}
{\pi \beta \over |k|} S(k;\beta) =
1 + \sum_{j=1}^\infty p_j(\beta / 2) \Big ( {|k| \over \pi \beta \rho}
\Big )^j, \quad |k| < {\rm min}
(2 \pi \rho, \pi \beta \rho)
\end{equation}
where $p_j(x)$ is a polynomial of degree $j$ which satisfies the functional
relation 
\begin{equation}\label{6.1}
p_j(1/x) = (-1)^j x^{-j} p_j(x).
\end{equation}
Equivalently, (\ref{6.1}) can be stated as requiring
\begin{eqnarray}\label{7.1}
p_j(x) & = & \sum_{l=0}^j a_{j,l} x^l, \qquad a_{j,l} =
a_{j,j-l} \quad (j \: {\rm even}) \\
\label{7.2}
p_j(x) & = & 
(x-1)\sum_{l=0}^{j-1} \tilde{a}_{j,l} x^l, \qquad \tilde{a}_{j,l} =
\tilde{a}_{j,j-1-l} \quad (j \: {\rm odd}).
\end{eqnarray}

Inspection of (\ref{4}) shows that the conjectured structure (\ref{6})
is correct at order $|k|$ and furthermore gives
\begin{equation}\label{6.2}
p_1(x) = (x-1),
\end{equation}
and thus $\tilde{a}_{1,0} = 1$ in (\ref{7.2}).
In Section 3 we use (\ref{Ha}) to verify that the structure (\ref{6})
is correct at order $k^2$ and we compute $p_2(x)$ explicitly. In Section
4 we use an exact an exact evaluation of the two-particle
distribution function for $\beta$ even \cite{Fo93aa} to rederive the
result of Section 3, and we also use this formula to
verify the structure (\ref{6}) at
order $k^4$ and to compute $p_4(x)$ explicitly.

Assuming the validity of (\ref{6}) we see that $p_j(x)$ can be computed
from knowledge of the coefficient of $|k|^j$ in $S(k;\beta)$, or the
coefficient of $|k|^j$ in $\partial^p S(k;\beta) / \partial \beta^p$
($p \le j$), for an appropriate number of distinct values of $\beta$.
Because the functional relation (\ref{5}) has via (\ref{7.1}) and (\ref{7.2})
been made a feature of (\ref{6}) the values of $\beta$ cannot be related
by $\beta \mapsto 4/\beta$. In Section 5 the known exact evaluation of
$S(k;\beta)$ to leading order in $\beta$ is reviewed, as are the 
exact evaluations of
$S(k;2)$ and $S(k;4)$. Also noted are the exact evaluations of $S(k,1)$
and $S(k;\beta)$ to leading order in $1/\beta$, which according to 
(\ref{5}) are related to $S(k;4)$ and $S(k;\beta)$ to leading order
in $\beta$ respectively.
All of these exact evaluations are in terms of
elementary functions, and so can be expanded to all orders in $k$. We
then present the exact evaluation
of $\partial S(k;\beta) / \partial \beta$ to leading
order in $\beta$, as well as the exact evaluation of
$\partial S(k;\beta) / \partial \beta$ evaluated at $\beta = 2$ and $\beta =4$.
The details of the latter two calculations are given in separate appendices.
Again the final expressions can be expanded to high order in $|k|$.
Using this data all polynomials in the expansion (\ref{6}) 
up to and including the term with $j=9$ can be computed. This expansion
is written out explicitly in the final section and 
some special features of the polynomials therein, relating to the sign of the
coefficients and the zeros, are noted. 
A physical interpretation of the functional equation, 
based on an analogy with a quantum many body system, which identifies an
equivalence between quasi-hole and quasi-particle states contributing to
$S(k;\beta)$ for $|k|$ small enough is given.
We end with some remarks on the
possible occurence of a functional equation analogous to (\ref{5})
in the two-dimensional one-component plasma.

\section{The functional equation}
\setcounter{equation}{0}
The Boltzmann factor (\ref{1}) also has the physical interpretation as the
absolute value squared of the exact ground state wave function, $|0\mg$
say, for the Calogero-Sutherland quantum many body Hamiltonian
\begin{equation}\label{cs}
H = - \sum_{j=1}^N{\partial^2 \over \partial x_j^2} +
\beta (\beta / 2 - 1) \Big ( {\pi \over L} \Big )^2
\sum_{1 \le j < k \le N}
{1 \over \sin^2 \pi (x_j - x_k)/L}.
\end{equation}
This Hamiltonian describes quantum particles on a circle of circumference
length $L$ interacting via the inverse square of the distance between the
particles. In the thermodynamic limit $N,L \to \infty$,
$N/L = \rho$ (fixed) the $N$ particle system becomes an infinite system on
a line with particle density $\rho$. The ground state dynamical 
density-density correlation function
\begin{equation}\label{h1}
\rho^{\rm dyn.}(0,x;t) := \ml 0 | n(0) e^{-i H t} n(x)
e^{i H t} | 0 \mg, \qquad
n(y) := \sum_{j=1}^N \delta(y-x_j)
\end{equation}
in the infinite system has been calculated exactly for all rational $\beta$
\cite{Ha95}. The fact that $(|0\mg)^2$ is proportional to (\ref{1}) tells us
that at $t=0$ (\ref{h1}) is equal to 
$$
\rho_{(2)}^T(0,x) + \rho \delta(x),
$$
where $\rho_{(2)}^T$ is the truncated two-body density,
for the log-gas system. Thus the exact evaluation of
$$
S(k,\beta) := \int_{-\infty}^\infty \Big (
\rho_{(2)}^T(0,x) + \rho \delta(x) \Big ) e^{i k x} \, dx
$$
for the log-gas follows from the exact evaluation of (\ref{h1}) for the
quantum system. Taking $\beta$ to be rational and setting
$$
\beta / 2 := p/q =: \lambda
$$
where $p$ and $q$ are relatively prime integers, the latter exact result
gives \cite{FJ97}
\begin{equation}\label{Ha}
S(k;\beta) = \pi C_{p,q}(\lambda) \prod_{i=1}^q
\int_0^\infty dx_i 
 \prod_{j=1}^p \int_0^1 dy_j Q_{p,q}^2
 F(q,p,\lambda|\{x_i,y_j\})\,
 \delta(k - Q_{p,q}),
\end{equation}
where
\begin{eqnarray}\label{C}
C_{p,q}(\lambda ) &:=& {\lambda^{2p(q-1)} \Gamma^2(p) \over 2 \pi^2 p! q!}
{\Gamma^q(\lambda) \Gamma^p(1/\lambda) \over 
\prod_{i=1}^q \Gamma^2(p-\lambda (i-1)) \prod_{j=1}^p \Gamma^2(1-(j-1)/\lambda)}
\nonumber \\
Q_{p,q} & := & 2 \pi \rho \Big ( \sum_{i=1}^q x_i + \sum_{j=1}^p y_j \Big )
\nonumber \\
F(q,p,\lambda|\{x_i,y_j\}) & := &
{\prod_{i<i'}|x_i - x_{i'}|^{2 \lambda}\prod_{j<j'}|y_j - y_{j'}|^{2 /\lambda}
\over 
\prod_{i=1}^q \prod_{j=1}^p (x_i + \lambda y_j)^{2}} \nonumber \\
&& \times {1 \over \prod_{i=1}^q (x_i(x_i+\lambda))^{1 - \lambda}
\prod_{j=1}^p(\lambda y_j(1 - y_j))^{1 - 1/\lambda}}
\end{eqnarray}

In the domain of integration of (\ref{Ha}) the integration variables are
all positive and because of the delta function are restricted to the
hyperplane
$$
\sum_{i=1}^q x_i + \sum_{j=1}^p y_j = {|k| \over 2 \pi \rho}.
$$
We see immediately from these constraints that the restriction $y_j < 1$
in the domain of integration is redundant for 
\begin{equation}\label{kl}
|k| < 2 \pi \rho.
\end{equation}
Thus assuming (\ref{kl}) we can extend the integration over $y_j$ to the
region $(0,\infty)$. Doing this and changing variables
$x_i \mapsto |k| x_i$ and $y_j \mapsto |k| y_j$ we see that for $|k|$
in the region (\ref{kl})
\begin{equation}\label{d2}
S(k;\beta) = \pi |k| C_{p,q}(\lambda)
\prod_{i=1}^q \int_0^\infty dx_i 
 \prod_{j=1}^p \int_0^\infty dy_j Q_{p,q}^2
 \hat{F}(q,p,\lambda|\{x_i,y_j\};k)\,
 \delta(1 - Q_{p,q}),
\end{equation}
where
\begin{eqnarray}\label{d2'}
\hat{F}(q,p,\lambda|\{x_i,y_j\};k)& = &
{1 \over \prod_{i=1}^q (x_i(1+kx_i/\lambda))^{1-\lambda}
\prod_{j=1}^p(y_j(1-ky_j))^{1-1/\lambda}} \nonumber \\
&& \times 
{\prod_{i<i'}|x_i - x_{i'}|^{2 \lambda}\prod_{j<j'}|y_j - y_{j'}|^{2 /\lambda}
\over 
\prod_{i=1}^q \prod_{j=1}^p (x_i + \lambda y_j)^{2}}
\end{eqnarray}
Notice that (\ref{d2'}) is such that the integral in (\ref{d2}) is analytic for
\begin{equation}\label{d2a}
|k| < {\rm min}\,(2\pi \rho, \pi \rho \beta).
\end{equation}
Thus according to the definition (\ref{4+}) we read off that
\begin{equation}\label{d3}
f(k;\beta) = 2 \pi^2 \lambda 
C_{p,q}(\lambda)
\prod_{i=1}^q \int_0^\infty dx_i 
 \prod_{j=1}^p \int_0^\infty dy_j Q_{p,q}^2
 \hat{F}(q,p,\lambda|\{x_i,y_j\};k)\,
 \delta(1 - Q_{p,q}).
\end{equation}

The functional equation (\ref{5}) is a simple consequence of this exact
formula. Thus we see that the integral in (\ref{d3}) is unchanged 
by the mapping
$\lambda \mapsto 1/\lambda$ (and thus $p \leftrightarrow q$) followed
by $k \mapsto - k/\lambda$. The precise functional equation (\ref{5})
follows provided we can show that
$$
C_{p,q}(\lambda) = \lambda^{2pq - 2} C_{q,p}(1/\lambda),
$$
which indeed readily follows from the definition of 
$C_{p,q}(\lambda)$ in (\ref{C}).

\section{Expanding $f(k;\beta)$ in terms of Dotsenko-Fateev type
integrals}
\setcounter{equation}{0}
Here we will develop a strategy based on the integral formula
(\ref{d3}) to expand $f(k,\beta)$ at order $k^2$. This relies on our
ability to compute certain generalizations of a limiting case of the
Dotsenko-Fateev integral. This same method has been used in
\cite{FZ96,FJ97} to compute the equivalent of $f(k,\beta)$ and its
derivative at $k=0$.

We first expand the integrand in (\ref{d3}) as a function of $k$.
According to (\ref{d2'}) we have
$$
\hat{F}(q,p,\lambda|\{x_i,y_j\};k) =
G(q,p,\lambda|\{x_i,y_j\}) \Big (
1 + \sum_{\nu=1}^\infty H_\nu(q,p,\lambda|\{x_i,y_j\}) k^\nu \Big )
$$
where
\begin{eqnarray}\label{u4}
G(q,p,\lambda|\{x_i,y_j\}) & = &
{\prod_{i<i'}|x_i - x_{i'}|^{2 \lambda}\prod_{j<j'}|y_j - y_{j'}|^{2 /\lambda}
\over 
\prod_{i=1}^q \prod_{j=1}^p (x_i + \lambda y_j)^{2}
\prod_{i=1}^q x_i^{1-\lambda} \prod_{j=1}^p y_j^{1-1/\lambda}} \nonumber \\
1 + \sum_{\nu = 1}^\infty
H_\nu(q,p,\lambda|\{x_i,y_j\}) k^\nu & = &
{1 \over \prod_{i=1}^q (1+kx_i/\lambda)^{1-\lambda}
\prod_{j=1}^p(1-ky_j)^{1-1/\lambda}}.
\end{eqnarray}
The coefficients $H_\nu$ are homogeneous polynomials in $\{x_i,y_j\}$ of
degree $\nu$.

Let us now introduce the notation
\begin{equation}\label{u5}
I_{p,q,\lambda}[h(\{x_i,y_j\})]
:=
\prod_{i=1}^q \int_0^\infty dx_i 
\prod_{j=1}^p \int_0^\infty dy_j \, Q_{p,q}^2
{G}(q,p,\lambda|\{x_i,y_j\})\,
\delta(1 - Q_{p,q}) h(\{x_i,y_j\}).
\end{equation}
Because of the presence of the delta function the value of $I_{p,q,\lambda}$
is unchanged if $Q_{p,q}^2$ is replaced by $Q_{p,q}^n$ for any $n$.
Doing this and also introducing the usual integral representation of the
delta function, we see by a change of variables as detailed in
\cite{FJ97} that for $h$ homogeneous of degree $\nu$
\begin{equation}\label{u6}
I_{p,q,\lambda}[h(\{x_i,y_j\})] =
{J_{p,q,\lambda,n}[h(\{x_i,y_j\})] \over (\nu + n -1)!} =
{J_{p,q,\lambda}[h(\{x_i,y_j\})] \over (\nu-1)!}
\end{equation}
where
$$
J_{p,q,\lambda,n}[h(\{x_i,y_j\})] :=
\prod_{i=1}^q \int_0^\infty dx_i 
\prod_{j=1}^p \int_0^\infty dy_j \, Q_{p,q}^n
{G}(q,p,\lambda|\{x_i,y_j\})\,
e^{-Q_{p,q}} h(\{x_i,y_j\})
$$
and $J_{p,q,\lambda} := J_{p,q,\lambda,0}$.

Recalling (\ref{d3}) and
(\ref{u4}) we see that in terms of the notation (\ref{u5})
\begin{equation}\label{u6f}
f(k;\beta) = C_{p,q}(\lambda) \Big (
I_{p,q,\lambda}[1] + \sum_{\nu=1}^\infty
I_{p,q,\lambda}[H_\nu(q,p,\lambda|\{x_i,y_j\})] k^\nu \Big ).
\end{equation}
The definition of $H_\nu$ in (\ref{u4}) shows
\begin{equation}\label{u6b}
H_2(q,p,\lambda|\{x_i,y_j\}) =
{(\lambda - 1)^2 \over 2 \lambda^2}
{Q_{p,q}^2 \over (2 \pi \rho)^2} - {\lambda - 1 \over 2 \lambda^2}
\Big ( \sum_{i=1}^q x_i^2 - \lambda \sum_{j=1}^p y_j^2 \Big ),
\end{equation}
so to compute $f(k,\beta)$ at order $k^2$ our task is to evaluate
\begin{equation}\label{u7}
I_{p,q,\lambda}[Q_{p,q}^2] \quad {\rm and} \quad
I_{p,q,\lambda}[ \sum_{i=1}^q x_i^2 - \lambda \sum_{j=1}^p y_j^2].
\end{equation}
Now because of the delta function in (\ref{u5})
\begin{equation}\label{u7'}
I_{p,q,\lambda}[Q_{p,q}^2]  =  I_{p,q,\lambda}[1],
\end{equation}
and we know from \cite{FJ97} that
\begin{equation}\label{u7b}
C_{p,q}(\lambda) I_{p,q,\lambda}[1] = 1.
\end{equation}
Thus our remaining task is to compute the second expression in (\ref{u7})
or equivalently, using (\ref{u6}), to compute
\begin{equation}\label{u7a}
J_{p,q,\lambda}[\sum_{i=1}^q x_i^2 - \lambda \sum_{j=1}^p y_j^2]
= q J_{p,q,\lambda}[x_i^2] - \lambda p J_{p,q,\lambda}[y_j^2]
\end{equation}
where the second equality, valid for any $1 \le i \le q$ and $1 \le j \le p$,
follows from the symmetry of the integrand.

For this purpose we first note formulas for $J_{p,q,\lambda}[h]$
in the cases $h=x_i^2$ and $h=y_j^2$. The formulas are
\begin{eqnarray}\label{dd1}
J_{p,q,\lambda}[x_i^2] & = &
{(2p - \lambda + 1) \over 2 \pi \rho} J_{p,q,\lambda}[x_i] -
{p \over \pi \rho} J_{p,q,\lambda} \Big [
{x_i^2 \over x_i + \lambda y_j} \Big ], \\
\label{dd2}
J_{p,q,\lambda}[y_j^2] & = &
{(2q - 1/\lambda + 1) \over  2 \pi \rho} J_{p,q,\lambda}[y_j]
- {\lambda q \over \pi \rho} J_{p,q,\lambda} \Big [
{x_j^2 \over x_i + \lambda y_j} \Big ].
\end{eqnarray}
The derivation of (\ref{dd1}) and (\ref{dd2}) uses a technique based on
the fundamental theorem of calculus. It was first used by Aomoto 
\cite{Ao87} in the context of the Selberg integral, and has been adapted
in \cite{FZ96} to the case of the Dotsenko-Fateev integral.

Let us give the details of the derivation of (\ref{dd1}) (the derivation
of (\ref{dd2}) is similar). From the definition (\ref{u4}) we see that
$$
{\partial \over \partial x_i} G(q,p,\lambda|\{x_i,y_j\})
= \Big ( {\lambda - 1 \over x_i} - 2 \sum_{j=1}^p
{1 \over x_i + \lambda y_j} + 2 \lambda
\sum_{i'=1 \atop i' \ne i}^q {1 \over x_i - x_{i'}} \Big )
G(q,p,\lambda|\{x_i,y_j\}).
$$
Thus
\begin{eqnarray}\label{r1}
0 & = & 
\prod_{i=1}^q \int_0^\infty dx_i 
\prod_{j=1}^p \int_0^\infty dy_j \, {\partial \over \partial x_i}
\Big ( x_i^2 
{G}(q,p,\lambda|\{x_i,y_j\})
e^{-Q_{p,q}} \Big ) \nonumber \\
& = & (\lambda + 1) J_{p,q,\lambda}[x_i] - 2 \sum_{j=1}^p
J_{p,q,\lambda} \Big [ {x_i^2 \over x_i + \lambda y_j} \Big ] +
2 \lambda \sum_{i'=1 \atop i' \ne i}^q 
J_{p,q,\lambda} \Big [ {x_i^2 \over x_i - x_{i'}} \Big ] - 2 \pi \rho
J_{p,q,\lambda}[x_i^2] \nonumber \\
& = & (\lambda + 1) J_{p,q,\lambda}[x_i] - 2p
J_{p,q,\lambda} \Big [ {x_i^2 \over x_i + \lambda y_j} \Big ]
+2 \lambda (q-1) 
J_{p,q,\lambda} \Big [ {x_i^2 \over x_i - x_{i'}} \Big ] - 2 \pi \rho
J_{p,q,\lambda}[x_i^2]  \nonumber \\
\end{eqnarray}
where the first equality follows from the fundamental theorem of calculus, while
the final equality, valid for any $j=1,\dots,p$ and any 
$i'=1,\dots,q$, ($i'\ne i$) follows by the symmetry of the integrand
with respect to $\{x_i\}$ and $\{y_j\}$. The symmetry of the integrand with
respect to $\{x_i\}$ also gives
$$
J_{p,q,\lambda} \Big [ {x_i^2 \over x_i - x_{i'}} \Big ]
= J_{p,q,\lambda} \Big [ {x_{i'}^2 \over x_{i'} - x_{i}} \Big ]
$$
so we have
$$
J_{p,q,\lambda} \Big [ {x_i^2 \over x_i - x_{i'}} \Big ]
= {1 \over 2} \bigg (
J_{p,q,\lambda} \Big [ {x_i^2 \over x_i - x_{i'}} \Big ] +
J_{p,q,\lambda} \Big [ {x_{i'}^2 \over x_{i'} - x_{i}} \Big ] \bigg )
= J_{p,q,\lambda}[x_i].
$$
Substituting in (\ref{r1}) implies (\ref{dd1}).

{}From (\ref{dd1}) and (\ref{dd2}) we see that
\begin{eqnarray}\label{u9}
\lefteqn{
q J_{p,q,\lambda}[x_i^2] - \lambda p J_{p,q,\lambda}[y_j^2]} \nonumber \\
&& = {q(2p - \lambda + 1) \over 2 \pi \rho}J_{p,q,\lambda}[x_i]
- {\lambda p (2q - 1/\lambda + 1) \over 2 \pi \rho}
J_{p,q,\lambda}[y_j] - {pq \over \pi \rho}
J_{p,q,\lambda} \Big [ {x_i^2 - \lambda^2 y_j^2 \over x_i + \lambda y_j}
\Big ] \nonumber \\
&& = {q(-\lambda +1) \over 2 \pi \rho}
J_{p,q,\lambda}[x_i] - {\lambda p (-1/\lambda + 1) \over 2 \pi \rho}
J_{p,q,\lambda}[y_j] \: = \:
- {\lambda - 1 \over (2 \pi \rho)^2} J_{p,q,\lambda}[Q_{p,q}]
\: = \:  - {\lambda - 1 \over (2 \pi \rho)^2} J_{p,q,\lambda}[1]
\nonumber \\
\end{eqnarray}

Recalling (\ref{u6b}), the results (\ref{u6}),
(\ref{u7'}), (\ref{u7a}) and (\ref{u9})
give that
$$
I_{p,q,\lambda}[H_2(q,p,\lambda|\{x_i\},\{y_j\}] =
{1 \over (2 \pi \rho)^2} {(\lambda - 1)^2 \over \lambda^2}
J_{p,q,\lambda}[1].
$$
Use of (\ref{u7b}) then gives that the term proportional to $k^2$ in
(\ref{u6f}) is equal to
\begin{equation}\label{u6g}
(\lambda - 1)^2 \Big ( {k \over 2 \pi \lambda \rho} \Big )^2 =
(\beta / 2 - 1)^2 \Big ( {k \over  \pi \beta \rho} \Big )^2.
\end{equation}
It follows from this that the structure (\ref{6}) is valid at order $k^2$
with
\begin{equation}\label{u6h}
p_2(x) = (x-1)^2.
\end{equation}

\section{Large-$x$ expansion of $\rho_{(2)}^T(0,x)$}
\setcounter{equation}{0}
We have already remarked that the large-$x$ expansion (\ref{2}) of the
charge-charge correlation, or what is the same thing for the one-component
log-gas, the large-$x$ expansion of $\rho_{(2)}^T(0,x)$, is equivalent to
the small-$k$ behaviour (\ref{3}) of $S(k;\beta)$. More generally the
expansion
\begin{equation}\label{e1}
\rho_{(2)}^T(0,x) \mathop{\sim}\limits_{x \to \infty}
\sum_{n=1}^\infty {c_n \over x^{2n}}
\end{equation}
is equivalent to the expansion
\begin{equation}\label{e2}
S(k;\beta) \mathop{\sim}\limits_{k \to 0} \pi
\sum_{n=1}^\infty {(-1)^n c_n \over (2n-1)!} |k|^{2n - 1}
\end{equation}
where the expansion (\ref{e2}) contains the terms singular in $k$
(i.e.~of odd order in $|k|$) only. This follows using the Fourier transform
$$
\int_{-\infty}^\infty {e^{ikx} \over x^{2n}} \, dx =
\pi {(-1)^n |k|^{2n-1} \over (2n - 1)!}
$$
from the theory of generalized functions (see e.g.~\cite{Li58}).

{}From the equivalence between (\ref{e1}) and (\ref{e2}) we see the fact,
following from (\ref{u6g}), that the term proportional to
$|k|^3$ in the small $k$ expansion of $S(k;\beta)$ is equal to
$$
\rho (\beta / 2 - 1)^2 \Big ( {|k| \over \pi \beta \rho} \Big )^3
$$
is equivalent to the statement that the term proportional to
$1/x^4$ in the large $x$ expansion of $\rho_{(2)}^T(0,x)$ is equal to
\begin{equation}\label{e3}
\rho^2 6 \beta (\beta/2 - 1)^2 \Big ( {1 \over \pi \beta \rho x}
\Big )^4.
\end{equation}
In this section we will derive (\ref{e3}) directly. We will also calculate
the $O(1/x^6)$ term and so explicitly determine the $O(|k|^5)$ term
in (\ref{e2}).

The starting point for our calculation is an exact $\beta$-dimensional
integral formula for the two-particle distribution
$\rho_{(2)}(0,x)$ valid for $\beta$ even. With
$$
S_n(a,b,c) := \prod_{j=0}^{n-1}{\Gamma(a+1+jc) \Gamma(b+1+jc)\Gamma(1+
(j+1)c) \over \Gamma(a+b+2+(N+j-1)c)\Gamma(1+c)}
$$
the formula gives that in the thermodynamic limit \cite{Fo93aa}
\begin{eqnarray}\label{i1}
\rho_{(2)}(0,x) & = & \rho^2 (\beta / 2)^\beta
{((\beta / 2)!)^3 \over \beta ! (3 \beta / 2)!}
{e^{-\pi i \beta \rho x} (2 \pi \rho x)^\beta \over
S_\beta(1-2/\beta,1-2/\beta,2/\beta)} \nonumber \\
&& \times \int_{[0,1]} du_1 \cdots du_\beta
\prod_{j=1}^\beta e^{2 \pi i \rho x u_j} u_j^{-1+2/\beta}
(1 - u_j)^{-1+2/\beta} \prod_{j < k}|u_k - u_j|^{4/\beta}.
\end{eqnarray}

In a previous analysis \cite{Fo93aa} it has been shown that the non-oscillatory
large-$x$ behaviour is determined by the integrand in the vicinity of the
endpoints 0 and 1, with the requirement that $\beta/2$ of the integration
variables are in the vicinity of the endpoint 0, while the remaining
$\beta/2$ integration variables are in the vicinity of the endpoint 1.
Thus we write $u_{\beta/2 + j} = 1 - v_j$ ($j=1,\dots, \beta/2$) 
(this introduces a combinatorial factor $\beta$ choose $\beta/2$
to account for the different ways of so partitioning the integration
variables) and then
expand the integrand (excluding the exponential
factors which involve $x$) in terms of the ``small'' variables $u_j,v_j$
($j=1,\dots,\beta/2$). In particular we must expand
\begin{equation}\label{i2}
\prod_{j=1}^{\beta/2}(1-u_j)^{-1+2/\beta}(1-v_j)^{-1+2/\beta}
\prod_{l,l'=1}^{\beta / 2}(1-u_l-v_{l'})^{4/\beta}.
\end{equation}
The function (\ref{i2}) is a symmetric function of the variables
$\{u_j\}$ and $\{v_j\}$ separately. Let $\{q_\kappa\}_\kappa$ be a polynomial
basis for symmetric functions with $\kappa$ denoting a partition (ordered
set of non-negative integers) of no more than $\beta/2$ parts, and suppose
furthermore that $q_\kappa$ is homogeneous of order $|\kappa|
:= \kappa_1 + \cdots + \kappa_{\beta / 2}$. Then we
can write
\begin{eqnarray}\label{i3}\lefteqn{
\prod_{j=1}^{\beta/2}(1-u_j)^{-1+2/\beta}(1-v_j)^{-1+2/\beta}
\prod_{l,l'=1}^{\beta / 2}(1-u_l-v_{l'})^{4/\beta}} \nonumber \\&&
= \sum_{\kappa, \mu} w_{\kappa, \mu}
q_\kappa(u_1,\dots,u_{\beta/2}) q_\mu(v_1,\dots,v_{\beta/2}).
\end{eqnarray}

Substituting (\ref{i3}) in (\ref{i1}), then following the procedure of
\cite{Fo93aa}, which involves extending the range of integration to
$u_j \in (0,\infty)$, $v_j \in (0,\infty)$ and changing variables
$u_j \mapsto 2 \pi i \rho x u_j$, $v_j \mapsto -2 \pi i \rho x v_j$
making use in the process of the fact that $q_\kappa$ is homogeneous
of degree $|\kappa|$, we obtain the non-oscillatory terms in the
large-$x$ asymptotic expansion of $\rho_{(2)}(0,x)$. This reads
\begin{eqnarray}\label{i4}
\rho_{(2)}(0,x) & \sim & \rho^2 \Big ( {\beta \atop \beta / 2} \Big )
(\beta / 2)^\beta
{((\beta / 2)!)^3 \over \beta ! (3 \beta / 2)!}
{1\over
S_\beta(1-2/\beta,1-2/\beta,2/\beta)} \nonumber \\ \times
&& \sum_{\kappa, \mu}
w_{\kappa, \mu} {K_{\beta, \kappa} K_{\beta, \mu} \over
i^{|\lambda| - |\mu|} (2 \pi \rho x)^{|\kappa| + |\mu|}}
\end{eqnarray}
where 
\begin{equation}\label{i5}
K_{\beta, \kappa} :=
\int_{[0,\infty)^{\beta / 2}} du_1 \cdots du_{\beta/2} \,
\prod_{l=1}^{\beta / 2} u_l^{-1 + 2/\beta} e^{-u_l}
\prod_{j < k} |u_k - u_j|^{4/\beta}
q_\kappa(u_1,\dots, u_{\beta/2}).
\end{equation}
The symmetry $w_{\kappa, \mu} = w_{\mu, \kappa}$ evident from (\ref{i3})
implies terms in (\ref{i4}) with $|\kappa| + |\mu|$ odd cancel. Therefore
the sum in (\ref{i4}) can be restricted to partitions such that
$|\kappa| + |\mu|$ is even, which means the asymptotic expansion only
contains inverse even powers of $x$.

To proceed further we must be able to compute the expansion coefficients
$w_{\kappa, \mu}$ as well as the integrals $K_{\beta, \kappa}$. For the
former task it is convenient to choose $q_\kappa$ equal to the monomial
symmetric polynomial $m_\kappa$, which is defined as the symmetrization of
the monomial $x_1^{\kappa_1} \cdots x_{\beta/2}^{\kappa_{\beta/2}}$ 
normalized so that the coefficient of 
$x_1^{\kappa_1} \cdots x_{\beta/2}^{\kappa_{\beta/2}}$  is unity.

First, we have the well known expansion
\begin{equation}\label{g1}
\prod_{j=1}^n(1-u_j)^a = \sum_{\ell(\kappa) \le n} a_\kappa
m_\kappa(u_1,\dots,u_n)
\end{equation}
where
\begin{equation}\label{g2}
a_\kappa = \prod_{p=1}^{\ell(\kappa)} a_{\kappa_p}, \qquad
a_k := {(-a)_k \over k!}
\end{equation}
with 
$\ell(\kappa)$ denoting the length of $\kappa$
(i.e.~number of non-zero parts). We can therefore
immediately expand the first product in (\ref{i3}) in terms of
monomial symmetric polynomials.

Consider next the expansion of the double product in (\ref{i3}). Making use
of the formulas
\begin{eqnarray}\label{g2a}
(1-x)^a & = & \sum_{n=0}^\infty {(-a)_n \over n!} x^n \\
\prod_{j=1}^n \Big ( \sum_{k=0}^\infty a_k t_j^k \Big ) & = &
\sum_{\ell(\kappa) \le n}
a_0^{N-\ell(\kappa)} a_\kappa m_\kappa(\{t_j\}) \label{g2b}
\end{eqnarray}
where $a_\kappa$ is specified by the first equality in (\ref{g2}), we see that
\begin{equation}\label{g3}
\prod_{j=1}^{\beta / 2} (1 - u_j - v)^{4/\beta} =
\sum_{\ell(\kappa) \le \beta / 2} (1 - v)^{2 - |\kappa|}
c_\kappa m_\kappa(u_1,\dots, u_{\beta / 2})
\end{equation}
where
$$
c_\kappa = \prod_{p=1}^{\ell(\kappa)} c_{\kappa_p}, \qquad
c_k := {(-4/\beta)_k \over k!}.
$$
Expanding the factor $(1-v)^{2 - |\kappa|}$ we can rewrite (\ref{g3}) as
$$
\prod_{j=1}^{\beta / 2} (1 - u_j - v)^{4/\beta} =
\sum_{n=0}^\infty w_n(u_1,\dots,u_{\beta/2};\beta) v^n
$$
for appropriate symmetric functions $w_n$. Replacing $v$ by $v_{j'}$
and forming the product over $j'$ using (\ref{g2b}) we obtain
$$
\prod_{j,j'=1}^{\beta / 2} (1 - u_j - v_{j'})^{4/\beta} =
\sum_{\ell(\kappa) \le \beta / 2}
w_0^{\beta/2 - \ell(\kappa)} w_\kappa m_\kappa(v_1,\dots,v_{\beta/2})
$$
where $w_\kappa := \prod_{p=1}^{\ell(\kappa)}
 w_{\kappa_p}$. The final step is to expand
$w_0^{\beta/2 - \ell(\kappa)} w_\kappa$ in terms of $\{m_\mu\}$ and so
obtain the expansion
\begin{equation}\label{g4}
\prod_{j,j'=1}^{\beta / 2} (1 - u_j - v_{j'})^{4/\beta} 
= \sum_{\mu, \kappa} t_{\mu,\kappa}
m_\mu(u_1,\dots,u_{\beta/2}) m_\kappa(v_1,\dots,v_{\beta/2}).
\end{equation}
The practical implementation of this procedure requires the use of computer
algebra. We work with arbitrary (positive integer) values of $\beta/2$.
Furthermore, we only include terms with $|\mu| + |\kappa| \le 6$ throughout
since according to (\ref{i4}) these terms suffice for the evaluation
of the coefficients of $1/x^{2n}$, $n \le 3$.

Having obtained the coefficients $t_{\mu,\kappa}$ in (\ref{g4}), we
multiply the series (\ref{g4}) with the two series of the form
(\ref{g1}) representing the first two products in (\ref{i2}),
expressing the answer in the form of (\ref{i3}),
and so determining the coefficients $w_{\kappa, \mu}$.
Again this step requires computer algebra.

With $w_{\kappa, \mu}$ in (\ref{i3}) determined, it remains to compute
the multiple integral (\ref{i5}) with $q_\mu = m_\mu$. For this task
we introduce a further basis of symmetric functions, namely the
Jack polynomials $\{P_\kappa^{(\beta/2)}(u_1,\dots,u_{\beta/2})\}$.
The Jack polynomials $P_\kappa^{(2/\beta)}(z_1,\dots,z_N)$ with
$z_j := e^{2\pi i x_j/L}$, when muliplied by the ground state wave
function $|0\mg$, are the eigenfunctions of the Calogero-Sutherland
Schr\"odinger operator (\ref{cs}) \cite{Fo94j}. Each polynomial
is homogeneous of degree $|\kappa|$ and has the expansion
\begin{equation}\label{g3a}
P_\kappa^{(\alpha)}(z_1,\dots,z_N) = m_\kappa +
\sum_{\mu < \kappa} a_{\kappa \mu} m_\mu
\end{equation}
where $<$ is the dominance partial ordering for partitions:
$\mu < \kappa$ if $|\kappa| = |\mu|$ with $\kappa \ne \mu$ and
$\sum_{i=1}^p \mu_i \le \sum_{i=1}^p \kappa_i$ for each
$p=1,\dots,N$. The coefficients $a_{\kappa \mu}$ can be calculated by
recurrence \cite{Ma95}. 

The significance of the Jack polynomial basis is that we have the explicit
integral evaluation
\begin{equation}\label{g3b}
{1 \over W_{a\alpha N}} \prod_{l=1}^N \int_0^\infty dt_l \,
t_l^a e^{-t_l} P_\kappa^{(\alpha)}(t_1,\dots,t_N)
\prod_{j<k}|t_k - t_j|^{2/\alpha}
=  P_\kappa^{(\alpha)}(1^N) [a + (N-1)/\alpha + 1]_\kappa^{(\alpha)},
\end{equation}
which is a limiting case of an integration formula due to Macdonald
\cite{Ma95}, Kadell \cite{Ka97g} and Kaneko \cite{Ka93}. In (\ref{g3b})
\begin{eqnarray*}
W_{a\alpha N} & = &
\prod_{l=1}^N \int_0^\infty dt_l \, t_l^a e^{-t_l}
\prod_{j < k} |t_k - t_j|^{2/\alpha} \: = \:
\prod_{j=0}^{N-1} {\Gamma(1 + (j+1)/\alpha) \Gamma(a+1+j/\alpha) \over
\Gamma(1+1/\alpha)}, \\
{}[u]_\kappa^{(\alpha)} & := &
\prod_{j=1}^N {\Gamma(u-(j-1)/\alpha + \kappa_j) \over
\Gamma(u - (j-1)/\alpha)}
\end{eqnarray*}
and $P_\kappa^{(\alpha)}(1^N)$ denotes $P_\kappa^{(\alpha)}(x_1,\dots,
x_N)$ evaluated at $x_1 = \cdots = x_N = 1$.

To make use of (\ref{g3b}) we must first express the monomial symmetric
polynomials $m_\kappa$ in terms of $\{P_\mu^{2/\beta}\}_{\mu \le \kappa}$,
which can be done using computer algebra
from knowledge of the expansion (\ref{g3a}). Substituting in (\ref{g3b})
allows the integrals $K_{\kappa \beta}$ to be computed.

After completing this procedure all terms in (\ref{i4}) for
$|\kappa| + |\mu| \le 6$ are known explicitly. Performing the sum and
simplifying we obtain
\begin{equation}\label{g6}
\rho_{(2)}(0,x) \sim \rho^2 \bigg (
1 - {1 \over \beta (\pi \rho x)^2} +
{3 (\beta - 2)^2 \over 2 \beta^3 (\pi \rho x)^4} -
{15 (\beta - 2)^2 (\beta^2 - 3\beta + 4) \over 2 \beta^5 (\pi \rho x)^6}
+ \cdots \bigg ).
\end{equation}
Note that this agrees with the known form (\ref{2}) for the term
$O(1/x^2)$, and the form (\ref{e3}) for the term $O(1/x^4)$.
The term $O(1/x^6)$, due to
the equivalence between (\ref{e1}) and (\ref{e2}), implies the
term $O(|k|^5)$ in the small-$k$ expansion of $S(k,\beta)$ is equal to
\begin{equation}\label{g6b}
(\beta/2-1)^2((\beta/2)^2 - {3 \over 2} (\beta/2) + 1) \Big ( {|k| \over 
\pi \beta} \Big )^5
\end{equation}
This is of the form of the conjecture (\ref{6}) with
\begin{equation}\label{g6c}
p_4(x) = (x-1)^2 (x^2 - {3 \over 2} x + 1).
\end{equation}

\section{$S(k;\beta)$
for special $\beta$}
\setcounter{equation}{0}
Let us assume the validity of (\ref{6}). The coefficients specifying
the polynomials $p_j(x)$ therein can be determined from knowledge of the
coefficient of $|k|^{j+1}$ in $S(k;\beta)$ or
$\partial^p S(k;\beta) / \partial \beta^p$ ($p \le j$) at special values
of $\beta$. Now in the context of random matrix theory $S(k;\beta)$ has
been evaluated in terms of elementary functions for $\beta = 1,2$ and 4.
The results are \cite{Me91}
\begin{eqnarray}
S(k;1) & = & \left \{ \begin{array}{ll}{|k| \over \pi} - {|k| \over 2 \pi}
\log \Big ( 1 + {|k| \over \pi \rho} \Big ),  & |k| \le 2 \pi \rho 
\\[.1cm]
2 - {|k| \over 2 \pi} \log  \Big ( 
{1 + |k|/\pi \rho \over -1 + |k|/\pi \rho} \Big ), & |k| \ge 2 \pi \rho 
\end{array} \right.
\label{er1} \\
S(k;2) & = & \left \{ \begin{array}{ll}{|k| \over 2\pi}, & |k| \le 2 \pi \rho 
\\[.1cm]
1, & |k| \ge 2 \pi \rho \end{array} \right.
\label{er2} \\
S(k;4) & = & \left \{ \begin{array}{ll}{|k| \over 4 \pi} - {|k| \over 8 \pi}
\log \Big | 1 - {|k| \over 2 \pi \rho} \Big |,  & |k| \le 4 \pi \rho 
\\[.1cm]
1, & |k| \ge 4 \pi \rho \end{array} \right.
 \label{er3}
\end{eqnarray}
Recalling the definition (\ref{4+}) of $f(k,\beta)$ we read off
\begin{eqnarray}
f(k;1) & = & 1 - {1 \over 2} \log \Big ( 1 + {k \over \pi \rho} \Big )
\label{er4} \\
f(k;2) & = & 1
\label{er5} \\
f(k;4) & = & 1 - {1 \over 2}  
\label{er6} \log \Big ( 1 - {k \over 2 \pi \rho} \Big )
\end{eqnarray}

The exact evaluation (\ref{er5}) implies that for all $j$ $p_j(x)$ contains
a factor of $(x-1)$. In the case of $j$ odd this gives no new information
since the factor $(x-1)$ was already deduced as a consequence of
the functional equation (\ref{6.1}). On the other hand, in the case $j$
even this fact together with the functional equation (\ref{6.1}) implies
\begin{equation}\label{pf}
p_j(x) = (x-1)^2 \sum_{l=0}^{j-2}b_{j,l} x^l, \qquad b_{j,l} = b_{j,j-2-l}
\quad (j \: {\rm even}).
\end{equation}

Consider now the constraints on the coefficients in (\ref{pf}) and (\ref{7.2})
which follow from (\ref{er4}) and (\ref{er6}). As (\ref{er4}) and (\ref{er6})
are related by the functional equation (\ref{5}), and this is built
into the structures (\ref{pf}) and (\ref{7.2}), only one of these exact
evaluations gives distinct information on $p_j(x)$. For definiteness
consider (\ref{er4}). We see that
\begin{equation}\label{pf1}
[k^j] f(k;1) = {1 \over 2} {(-1)^{j} \over j (\pi \rho)^j}, \quad j \ge 1
\end{equation}
where the notation $[k^j]$ denotes the coefficient of $k^j$.
Recalling (\ref{6}), (\ref{pf}) and (\ref{7.2}) this implies, for 
$j$ even,
\begin{equation}
{1 \over j}  =  {1 \over 2}\Big ( (1+2^{-(j-2)})b_{j,0}
+ (2^{-1} + 2^{-(j-3)})b_{j,1} + \cdots + (2^{-j/2 +2} + 2^{-j/2})b_{j,j/2-2}
+ 2^{-j/2 +1} b_{j,j/2-1} \Big ),  \label{nt1}
\end{equation}
while for $j$ odd
\begin{eqnarray}
{1 \over j}  & =  & \Big ( (1+2^{-(j-1)})\tilde{a}_{j,0}
+ (2^{-1} + 2^{-(j-2)})\tilde{a}_{j,1} + 
\cdots + (2^{-(j-1)/2 +1} + 2^{-(j-1)/2-1})\tilde{a}_{j,(j-1)/2-1}
\nonumber \\
&&
+ 2^{-(j-1)/2} \tilde{a}_{j,(j-1)/2} \Big ). 
\label{nt2}
\end{eqnarray}
In the case $j=1$ (\ref{nt2}) gives $\tilde{a}_{j,0}=1$ which reclaims
(\ref{6.2}), while 
in the case $j=2$ (\ref{nt1}) gives $b_{j,0}=1$ which reclaims
(\ref{u6h}).

The exact form of $S(k;\beta)$ in the weak coupling scaling limit
$\beta \to 0$, $k \to 0$, $k/\beta$ fixed is also available. Introducing
the dimensionless Fourier transforms
$$
\tilde{S}(k;\beta) := \rho \int_{-\infty}^\infty
\Big ( \rho_{(2)}^T(0,x) + \rho \delta(x) \Big ) e^{i \rho x k} \, dx,
\qquad \tilde{\Phi}(k) := \rho \int_{-\infty}^\infty \Phi(x)
e^{i \rho x k} \, dx
$$
where 
$\Phi(x) := - \log|x|$ is the pair potential of the log-gas (thus the
integral in the definition of the $\tilde{\Phi}(k)$ is to be interpreted as
a generalized function) we have \cite{HM90}
\begin{equation}\label{5.30a}
\tilde{S}(k;\beta) \sim 1 - {\beta \tilde{\Phi}(k) \over 1 +
\beta \tilde{\Phi}(k)}.
\end{equation}
Since 
\begin{equation}\label{fte}
\tilde{\Phi}(k) = {\pi \over |k|},
\end{equation}
and noting $\tilde{S}(k;\beta) = S(k\rho;\beta)/\rho$
we thus have that in the weak coupling
scaling limit
\begin{equation}\label{us}
S(k,\beta) = \rho \Big ( 1 - {1 \over 1 + |k|/\pi \beta \rho} \Big ).
\end{equation}
Expanding (\ref{us}) in the form (\ref{6}) and recalling (\ref{pf})
and (\ref{7.2}) we deduce
\begin{equation}\label{us1}
\tilde{a}_{j,0} = 1 \qquad {\rm and} \qquad b_{j,0}=1
\end{equation}
for all $j$. Using (\ref{us1}) in (\ref{nt1}) and (\ref{nt2}) gives that
in the case $j=3$, $\tilde{a}_{j,1} = - {11 \over 6}$, and in the case $j=4$,
$b_{j,1} = - {3 \over 2}$. The latter result reclaims (\ref{g6b})
while the former result together with (\ref{us1}) gives
\begin{equation}\label{p3}
p_3(x) = (x-1)(1 - {11 \over 6} x + x^2 ).
\end{equation}

An alternative way to derive (\ref{us1}) is to consider the $\beta \to \infty$
low temperature limit. In this limit the system behaves like an harmonic 
crystal, for which we have available the analytic formula
\cite{Fo93}\footnote{The denominator of the exponent in (3.10) of
\cite{Fo93} contains a spurious factor of $\pi^2$ which is corrected in
(\ref{har})}
\begin{equation}\label{har}
\rho_{(2)}^{\rm (har)}(x;0) = \rho^2 \sum_{p=-\infty \atop p \ne 0}^\infty
\Big ( {\beta \over 4 \pi f(p)} \Big )^{1/2}
e^{-\beta(p-\rho x)^2/4f(p)}
\end{equation}
where
$$
f(p) = {1 \over \pi^2} \int_0^{1/2} {1 - \cos 2 \pi p t \over t - t^2}
\, dt.
$$
Taking the Fourier transform gives for $|k| < 2 \pi \rho$
\begin{eqnarray}
S^{\rm (har)}(k;\beta) & = & \rho \sum_{p=-\infty}^\infty \Big (
e^{-k^2 f(p) / \beta \rho^2} - 1 \Big ) e^{ikp/\rho} \nonumber \\
& \mathop{\sim}\limits_{\beta \to \infty} & -
\rho {k^2 \over \beta \rho^2} \sum_{p=-\infty}^\infty f(p)
e^{ikp/\rho}  \: = \: {|k|/\pi \beta \over 1 - |k|/2\pi \rho}.
\end{eqnarray}
This formula maps to the weak coupling result (\ref{5.30a}) under the
action of the functional equation (\ref{5}) and so implies (\ref{us1}).

\section{Perturbation about $\beta = 0$}
\setcounter{equation}{0}
The formula (\ref{us}) is just the first term in a systematic weak
coupling renormalized Mayer series expansion in $\beta$. In the case of
the two-dimensional one-component plasma, low order terms of this
expansion have recently been analyzed by Kalinay et al.~\cite{KMST99}.
Results from that study can readily be transcribed to the case of the
one-component log-gas.

Formally, the renormalized Mayer series expansion is for the
dimensionless free energy $\beta \bar{F}^{\rm ex}$ (in \cite{KMST99}
our $\beta \bar{F}^{\rm ex}$ is written $-\beta \bar{F}^{\rm ex}$), and
one computes the direct correlation function via the 
functional differentiation formula
\begin{equation}\label{o1}
c(0,x) = - {\delta^2 (\beta \bar{F}^{\rm ex}) \over \delta \rho_{(1)}(0)
\delta \rho_{(1)}(x)}.
\end{equation}
The Ornstein-Zernicke relation gives that the dimensionless Fourier transform
of the direct correlation function, $\tilde{c}(k,\beta)$ say, is related to
the dimensionless structure function $\tilde{S}(k;\beta)$ by
\begin{equation}\label{o2}
\tilde{c}(k;\beta) = 1 - {1 \over \tilde{S}(k;\beta)}
\end{equation}
so expanding $\tilde{c}(k,\beta)$ about $\beta = 0$ with $k/\beta$ fixed is
equivalent to expanding $\tilde{S}(k;\beta)$ about $\beta = 0$ with 
$k/\beta$ fixed.

Now, transcribing the results of \cite{KMST99} we read off that the
weak coupling diagrammatic expansion of $c(x_1,x_2)$ starts as
\begin{equation}
c(x_1,x_2) = -\beta \Phi(x_1,x_2) + {1 \over 2!} \Big (K(x_1,x_2) \Big )^2 
+ \cdots
\end{equation}
where
\begin{equation}
K(x_1,x_2)  = -\beta \pi \int_{-\infty}^\infty
{e^{i k (x_1 - x_2)} \over |k| + \kappa} \, dk 
\end{equation}
with $\kappa = \beta \pi \rho$.
This implies
\begin{equation}
\tilde{c}(k;\beta) = - {\beta \pi \over |k|} + {1 \over 2} \rho
\int_{-\infty}^\infty {dl \over 2 \pi} \, {\beta \pi \over |l| +
\kappa} {\beta \pi \over |\rho k - l| + \kappa}.
\end{equation}
The integral is straightforward (consider separately the ranges of $l$ such 
that $l >0$ ($l <0$) and $\rho k-l >0$ ($\rho k - l <0$)). In terms of
$k' := \rho k / \kappa = k/\pi \beta$,
\begin{equation}
\tilde{c}(k;\beta) =
- {1 \over |k'|} + \beta {1 + |k'| \over |k'|(2 + |k'|)}
\log(1 + |k'|) + O(\beta^2),
\end{equation}
or equivalently using (\ref{o2})
\begin{equation}\label{re1}
S(k;\beta) = \rho {|k/\kappa| \over 1 + |k/\kappa|} +
\beta \rho {|k/\kappa| \over (1 + |k/\kappa|)(2+|k/\kappa|)}
\log(1 + |k/\kappa|) + O(\beta^2).
\end{equation}
Notice that the leading order term in (\ref{re1}) reproduces (\ref{5.30a}).

The exact result (\ref{re1}) gives the explicit value of the coefficient
of $x$ in the polynomial $p_j(x)$. Thus recalling (\ref{7.2}) and
(\ref{pf}) we have
\begin{eqnarray}\label{re2}
{1 \over 2}(b_{j,1} - 2) & = & [x^j] {1 \over (1+x)(2+x)}
\log (1+x), \qquad (j \quad {\rm even}) \nonumber \\
{1 \over 2} (1 - \tilde{a}_{j,1}) & = & [x^j] {1 \over (1+x)(2+x)}
\log (1+x), \qquad (j \quad {\rm odd}).
\end{eqnarray}
Furthermore, a simple calculation gives
\begin{equation}\label{re3}
[x^j] {1 \over (1+x)(2+x)}
\log (1+x) = (-1)^j \sum_{q=1}^j {1 \over q}(1 - 2^{q-j})
\end{equation}
so we have for example
\begin{equation}\label{re4}
\tilde{a}_{5,1} = - {91 \over 30}, \quad
b_{6,1} = - {31 \over 15}, \quad
\tilde{a}_{7,1} = - {1607 \over 420}, \quad
b_{8,1} = - {263 \over 84}, \quad
\tilde{a}_{9,1} = - {791 \over 180}.
\end{equation}

Substituting $\tilde{a}_{5,1}$ from (\ref{re4}) and 
$\tilde{a}_{5,0}$ from
(\ref{us1}) in (\ref{nt1}) shows $\tilde{a}_{5,2} = {62 \over 15}$.
Similarly, the value of $b_{6,1}$ above allows us to deduce
that $b_{6,2} = {13 \over 4}$. Thus we have
\begin{eqnarray}\label{re4'}
p_5(x) & = & (x-1)(x^4 - {91 \over 30} x^3 + {62 \over 15} x^2
- {91 \over 30} x + 1) \nonumber \\
p_6(x) & = & (x-1)^2(x^4 - {37 \over 15} x^3 + {13 \over 4} x^2
- {37 \over 15} x + 1).
\end{eqnarray}

We remark that according to the conjecture (\ref{6}), the expansion of
$S(k,\beta)$ about $\beta = 0$ should have the structure
\begin{equation}\label{re5}
S(k,\beta) = f_0(k/\kappa) + \beta f_1(k/\kappa) + \beta^2
f_2(k/\kappa) + \cdots
\end{equation}
where
\begin{equation}\label{re6}
f_j(u) = u^j ( c_{j,0} + c_{j,1} u + \cdots ).
\end{equation}
Consideration of the analysis of \cite{KMST99} reveals that the
structure (\ref{re5}) will indeed result from the weak coupling
expansion, however the structure (\ref{re6}) is not immediately evident.
(Of course the explicit form $f_2$ as revealed by (\ref{re1}) exhibits
this structure.) 

\section{Perturbation about $\beta = 2$ and $\beta = 4$}
\setcounter{equation}{0}
A feature of the couplings $\beta=1,2$ and 4 is that the $n$-particle
distribution functions are known for each $n=2,3,\dots$
\cite{Me91}. Introducing the
dimensionless distribution
$$
g(x_1,\dots,x_n) := \rho_{(n)}(x_1,\dots,x_n) / \rho^n
$$
we can use our knowledge of $g(x_1,\dots,x_n)$ for $n=2,3$ and 4 at these
specific $\beta$ to expand $g(x_1,x_2)$ about $\beta = \beta_0$
to first order in $\beta - \beta_0$. Thus with $\Phi(x_1,x_2) :=
- \log |x_1 - x_2|$ we have \cite{Ja81}
\begin{eqnarray}\label{ny0}
g(x_1,x_2;\beta) & = & g(x_1,x_2) + (\beta - \beta_0) \bigg \{
- g(x_1,x_2) \Phi(x_1,x_2) \nonumber \\
&& - 2 \rho \int_{-\infty}^\infty
\Big ( g(x_1,x_2,x_3) - g(x_1,x_2) \Big ) \Phi(x_1,x_3) \, dx_3 \nonumber \\
&& - {1 \over 2} \rho^2 \int_{-\infty}^\infty \Big (
g(x_1,x_2,x_3,x_4) - g(x_1,x_2) g(x_3,x_4) - g(x_1,x_2,x_3)
\nonumber \\
&& - g(x_1,x_2,x_4) + 2 g(x_1,x_2) \Big ) \Phi(x_3,x_4) \, dx_3 dx_4
\bigg \} + O((\beta - \beta_0)^2)
\end{eqnarray}
where on the right hand side the dimensionless distributions are evaluated 
at $\beta = \beta_0$. Here we will compute this first order correction,
and the corresponding first order correction for $S(k;\beta)$, in the cases
$\beta_0=2$ and $\beta_0 = 4$ (we do not consider $\beta_0=1$ because
of its relation to  $\beta_0=4$ via the functional equation (\ref{5})).  

Now, in the case $\beta_0 = 2$ we have 
\begin{equation}\label{ny1}
g(x_1,\dots,x_n) = \det \Big [ P_2(x_j,x_k) \Big ]_{j,k=1,\dots,n},
\qquad
P_2(x,y) := {\sin \pi \rho (x-y) \over \pi \rho (x-y)}
\end{equation}
while in the case $\beta_0=4$
\begin{equation}\label{ny2}
g(x_1,\dots,x_n) = {\rm qdet} \Big [ P_4(x_j,x_k) \Big ]_{j,k=1,\dots,n}
\end{equation}
where
\begin{equation}\label{ny2'}
P_4(x_j,x_k) = \left [ \begin{array}{cc} \displaystyle
{\sin 2 \pi \rho x_{jk} \over 2 \pi \rho x_{jk}} & {\rm Si}\, 
( 2 \pi \rho x_{jk}) \\[.2cm]
\displaystyle {1 \over 2 \pi \rho} {d \over d x_{jk}}
\Big ( {\sin 2 \pi \rho x_{jk} \over 2 \pi \rho x_{jk}} \Big ) &
\displaystyle
{\sin 2 \pi \rho x_{jk} \over 2 \pi \rho x_{jk}} \end{array} \right ]
\end{equation}
with $x_{jk} := x_j - x_k$ and Si$(x)$ denoting the complimentary
sine integral, defined in terms of the sine integral si$(x)$ by
\begin{equation}\label{si}
{\rm Si}\,(x) = \int_0^x {\sin t \over t} \, dt =
{\pi \over 2} + {\rm si}\,(x), \quad
 {\rm si}\,(x) := - \int_x^\infty {\sin t \over t} \, dt.
\end{equation}
In (\ref{ny2}) qdet denotes quaternion determinant, which can be defined as
\begin{equation}\label{qdet}
{\rm qdet} \Big [ P_4(x_j,x_k) \Big ]_{j,k=1,\dots,n} =
\sum_{P \in S_n} (-1)^{n-l} \prod_1^l\Big ( P_4(x_a,x_b) P_4(x_b,x_c)
\cdots P_4(x_d,x_a) \Big )^{(0)}
\end{equation}
where the superscript $(0)$ denotes the operation ${1 \over 2} {\rm Tr}$,
$P$ is any permutation of the indicies $(1,\dots,n)$ consisting of $l$
exclusive cycles of the form $(a \to b \to c \cdots \to d \to a)$ and
$(-1)^{n-l}$ is equal to the parity of $P$. Note that this reproduces the
definition of an ordinary determinant in the case that $P_4$ is a
multiple of the identity.

The task now is to substitute (\ref{ny1}) in the case $\beta_0=2$ and
(\ref{ny2}) in the case $\beta_0=4$, and to compute the integrals. Consider
first the case $\beta_0=2$. After some calculation (see Appendix A) we find
\begin{eqnarray}\label{af0}
g(0,x;\beta) & = & 1 - \Big ( {\sin \pi \rho x \over \pi \rho x} \Big )^2
+ (\beta - 2) \bigg \{ {1 \over 2}
\Big ( {\sin \pi \rho x \over \pi \rho x} \Big )^2
- {\sin 2 \pi \rho x \over 2 \pi \rho x} + {\rm ci}\, (2 \pi \rho x) \nonumber \\
&& + {1 \over 2 (\pi \rho x)^2} \Big (
(\log 2 \pi \rho |x| + C) \cos 2 \pi \rho x - {\rm ci}\,(2 \pi \rho x)
\Big ) \bigg \} + O((\beta - 2)^2)
\end{eqnarray}
where $C$ denotes Euler's constant while
\begin{equation}\label{ci}
{\rm ci}(x) = C + \log |x| + \int_0^x {\cos t - 1 \over t} \, dt =
- \int_x^\infty {\cos t \over t} \, dt
\end{equation}
denotes the cosine integral. From this we can compute (again see Appendix
A) that up to terms $O((\beta - 2)^2)$ 
\begin{equation}\label{af1}
S(k;\beta)  =  \left \{ \begin{array}{ll}
 {|k| \over 2 \pi} + (\beta - 2) \rho \bigg \{
{1 \over 2} \log \Big ( 1 - {k^2 \over (2 \pi \rho)^2} \Big ) +
{|k| \over 4 \pi \rho} \log {2 \pi \rho + |k| \over 2 \pi \rho - |k|}
- {|k| \over 4 \pi \rho} \bigg \}, & 
|k| < 2 \pi \rho, \\
\rho + (\beta - 2) \rho \bigg \{
{1 \over 2}  \log {|k| + 2 \pi \rho \over  |k| - 2 \pi \rho}
+ {|k| \over 4 \pi \rho} \log \Big ( 1 - {(2 \pi \rho)^2 \over
k^2} \Big )
- {\pi \rho \over |k|} \bigg \}, & 
|k| > 2 \pi \rho. \end{array} \right.
\label{af2}
\end{equation}

Let us consider the consequence of (\ref{af1}) in regards to the expansion
(\ref{6}). For $|k| < 2 \pi \rho$
we observe that all terms but the one proportional to $|k|$
are even in $k$. This is consistent with $p_j(x)$ having the quadratic
factor $(x-1)^2$ for $j$ odd (recall (\ref{pf})), but only a linear
factor for $j$ even (recall (\ref{7.2})). Moreover, we can use
(\ref{af1}) to derive a linear equation for the coefficients
$\{\tilde{a}_j\}$. First we differentiate (\ref{af1}) with respect to
$\beta$, set $\beta = 2$ and expand about $k=0$ to obtain
$$
{\partial S(k;\beta) \over \partial \beta} \Big |_{\beta = 2}
= - {1 \over 2} {|k| \over 2 \pi \rho}
+ \sum_{j=1}^\infty {1 \over 2j (2j - 1)}
\Big ( {|k| \over 2 \pi \rho} \Big )^{2j}, \quad |k| < 2 \pi \rho.
$$
Recalling (\ref{6}) and (\ref{pf}) this in turn implies
\begin{equation}\label{cw}
{1 \over 2j(2j-1)} = {1 \over 2} \Big ( 2 \tilde{a}_{2j-1, 0} +
2  \tilde{a}_{2j-1,1} + 
\cdots + 2  \tilde{a}_{2j-1,j-2} +  \tilde{a}_{2j-1,j-1} \Big ).
\end{equation}
In the case $j=4$ we deduce from this equation, (\ref{us1}), (\ref{re4})
and (\ref{nt2}) that
\begin{equation}\label{p7}
p_7(x) = (x-1) \Big ( 1 - {1607 \over 420} x +
{2011 \over 280} x^2 - {911 \over 105} x^3 + {2011 \over 280} x^4
- {1607 \over 420} x^5 + x^6 \Big ).
\end{equation}

Consider now the case $\beta_0=4$. Due to $P_4$ in (\ref{ny2})
being a $2 \times 2$ matrix, the calculation required to compute (\ref{ny0})
is more lengthy and tedius than in the case $\beta_0=2$, although the
the common structure of $n$-point distributions means the two cases are
analogous. Some details are given in Appendix B. Our final expression
for $g(x_1,x_2;\beta)$ is given by (\ref{Bg}). We find its Fourier transform
can be computed explicitly in terms of elementary functions, together with
the dilogarithm
\begin{equation}\label{dilog}
{\rm dilog}\, (x) := \int_1^x {\log t \over 1 - t} \, dt.
\end{equation}
Explicitly, with $\rho = 1$ for notational convenience, up to terms
$O((\beta - 4)^2)$
\begin{equation}\label{b4}
S(k,\beta) = S(k,4) + (\beta - 4) \Big ( - {\pi \over |k|} +
\hat{B}_0 (k) + 2 \hat{B}_1(k) - 4 \hat{B}_3(k) + 
2 \hat{B}_5(k) + \hat{B}_6(k) - \hat{B}_7(k) \Big )
\end{equation}
where
\begin{eqnarray}\label{B0}
\hat{B}_0(k) & = & - {3 \over 2} + {3|k| \over 8 \pi} +
{|k| \over 4 \pi} \log \Big ( {4 \pi + |k| \over |k|} \Big )
+ \Big ( C + {1 \over 2} \log(16 \pi^2 - k^2) \Big ) \Big ( 1 -
{|k| \over 4 \pi} + {|k| \over 8 \pi} \log \Big | 1 -
{|k| \over 2 \pi} \Big | \Big ) \nonumber \\
&& + {|k| \over 16 \pi} \bigg ( {\rm dilog} \, \Big (
{|k| \over 2 \pi + |k|} \Big ) -
{\rm dilog} \, \Big ( {4 \pi + |k| \over 2 \pi + |k|} \Big ) -
\log \Big | 1 - {|k| \over 2 \pi} \Big |
\log \Big ( {4 \pi + |k| \over |k|} \Big ) + g_1(k) \bigg ) \nonumber \\
&& + {2 \pi - |k| \over 8 \pi} \log \Big | 1 - {|k| \over 2 \pi} \Big |,
\qquad |k| < 4 \pi,
\end{eqnarray}
\begin{eqnarray}
\hat{B}_0(k) & = &
{1 \over 2} \log \Big ( {|k| + 4 \pi \over |k| - 4 \pi} \Big ) 
+ {|k| \over 8 \pi} \log \Big ( {k^2 - 16 \pi^2 \over k^2} \Big )
+ {|k| \over 16 \pi} \bigg ( {\rm dilog} \,
\Big ( {|k| \over |k| + 2 \pi} \Big ) \nonumber \\ &&
+
{\rm dilog} \,
\Big ( {|k| \over |k| - 2 \pi} \Big ) - {\rm dilog} \,
\Big ( {|k| + 4 \pi \over |k| + 2 \pi} \Big ) -
{\rm dilog} \,
\Big ( {|k| - 4 \pi \over |k| - 2 \pi} \Big ) \bigg ), \qquad |k| > 4 \pi,
\end{eqnarray}
\begin{equation}
\hat{B}_1(k) = \left \{\begin{array}{ll}
{\pi \over |k|} \Big ( 1 - {|k| \over 4 \pi} + {|k| \over 8 \pi}
\log \Big | 1 - {|k| \over 2 \pi} \Big | \Big ), & |k| < 4 \pi \\[.2cm]
0, & |k| > 4 \pi, \end{array} \right.
\end{equation}
\begin{eqnarray}
\hat{B}_3(k) & = & - {3 \over 2} + {3 |k| \over 8 \pi} +
C \Big ( 1 - {|k| \over 4 \pi} + {|k| \over 8 \pi} \log \Big |
1 - {|k| \over 2 \pi} \Big | \Big )
+ \Big ( {1 \over 8} - {3|k| \over 32 \pi} \Big )
\log \Big |
1 - {|k| \over 2 \pi} \Big | \nonumber \\&&
+ {|k| \over 64 \pi} \Big ( \log \Big | 1 - {|k| \over 2 \pi} \Big |
\Big )^2 + {1 \over 8 \pi} (4 \pi - |k|)
\log (4 \pi - |k|) - {1 \over 8 \pi} |k| \log |k| +
{1 \over 2} \log 4 \pi \nonumber \\
&& + {|k| \over 32 \pi} \bigg (
{\rm dilog} \, \Big ( {|k| \over 2 \pi + |k|} \Big ) + {\pi^2 \over 12}
- {\rm dilog} \,\Big ({4 \pi \over 2 \pi + |k|} \Big )
- {\rm dilog} \, \Big ( {|k| \over 2 \pi} \Big ) -
{\rm dilog} \, \Big ({4 \pi - |k| \over 2 \pi} \Big ) \nonumber \\&&
+2 \log(2 \pi) \log \Big | 1 - {|k| \over 2 \pi} \Big |
+ \log \Big (2 \pi + |k| \Big ) \log \Big | 1 - {|k| \over 2 \pi} \Big |
+ g_2(k) \bigg ), \quad |k| < 4 \pi, \nonumber \\
\hat{B}_3(k) &  = & 0, \qquad |k| > 4 \pi,
\end{eqnarray}
\begin{equation}
\hat{B}_5(k)  =  \left \{ \begin{array}{ll}
\hat{B}_3(k) - {|k| \over 128 \pi} \Big ( \log \Big | 
1 - {|k| \over 2 \pi} \Big |
\Big )^2 +  {|k| \over 32 \pi} g_3(k), & |k| < 4 \pi  \\
0, & |k| > 4 \pi \end{array} \right.
\end{equation}
\begin{eqnarray}
\hat{B}_6(k) & = & - {3 \over 2} + {3|k| \over 8 \pi} - {|k| \over 16 \pi}
\log \Big | 1 - {|k| \over 2 \pi} \Big | +
\Big ( C + \log (4 \pi - |k|) \Big ) \Big ( 1 - {|k| \over 4 \pi}
+ {|k| \over 8 \pi} \log \Big | 1 - {|k| \over 2 \pi} \Big | \Big )
\nonumber \\
&& + {|k| \over 32 \pi}\bigg ( {\pi^2 \over 3} -
{\rm dilog} \, \Big ( {|k| \over 2 \pi} \Big )  -
\log \Big | 1 - {|k| \over 2 \pi} \Big | \log \Big ( {|k| \over 2 \pi}
\Big ) - 2 {\rm dilog} \,\Big ( {4 \pi - |k| \over 2 \pi} \Big )
\nonumber \\ &&
- 2 \log \Big | 1 - {|k| \over 2 \pi} \Big |
\log \Big ( {4 \pi - |k| \over 2 \pi} \Big ) + g_4(k) \bigg ), \quad
|k| < 4 \pi, \nonumber \\
\hat{B}_6(k) & = & 0,  \quad |k| > 4 \pi,
\end{eqnarray}
\begin{equation}\label{B7}
\hat{B}_7(k) = \left \{ \begin{array}{ll}
{\pi \over |k|} \Big ( 1 - {|k| \over 4 \pi}
+ {|k| \over 8 \pi} \log \Big | 1 - {|k| \over 2 \pi} \Big | \Big )^2, &
|k| < 4 \pi \\
0, & |k| > 4 \pi, \end{array} \right.
\end{equation}
with
\begin{eqnarray}\label{gg}
g_1(k) & = & \left \{ \begin{array}{ll}
{\rm dilog} \Big ( {4 \pi - |k| \over 2 \pi - |k|} \Big ) -
{\rm dilog} \Big ( {2 \pi \over 2 \pi - |k|} \Big ) - {\pi^2 \over 6}
- \log \Big ( 1 - {|k| \over 2 \pi} \Big ) \log \Big ( {4 \pi - |k| \over 
2 \pi - |k|} \Big ), & |k| < 2 \pi \\[.2cm]
{\rm dilog} \Big ( {|k| \over |k| - 2 \pi} \Big ) -
{\rm dilog} \Big ( {2 \pi \over |k| - 2 \pi} \Big ) - {\pi^2 \over 6}
+ \log \Big ( {2 \pi \over |k| - 2\pi} \Big ) \log \Big (
{|k| \over |k| - 2 \pi} \Big ), & 2 \pi < |k| < 4 \pi, \end{array}
\right. \nonumber \\
g_2(k) & = & \left \{ \begin{array}{ll}
{\rm dilog} \Big ( {4 \pi - |k| \over 2 \pi - |k|} \Big ) - {\pi^2 \over 6}
+ \log(2 \pi - |k|) \log \Big (1 - {|k| \over 2 \pi} \Big ),
& |k| < 2 \pi \\[.2cm]
- {\rm dilog} \Big ( {2 \pi \over |k| - 2 \pi} \Big ) +
\log(4 \pi - |k|) \log \Big ( {|k| \over 2 \pi} - 1 \Big ),
& 2 \pi < |k| < 4 \pi \end{array} \right.
\nonumber \\
g_3(k) & = & \left \{ \begin{array}{ll}
{1 \over 2} \Big ( {\pi^2 \over 6} - 2 {\rm dilog} \Big (
{2 \pi \over |k|} \Big ) -
\log \Big (
{2 \pi \over |k|} \Big ) \log \Big ( {(2 \pi - |k|)^2 \over 2 \pi |k|}
\Big ) \Big ), & |k| < 2 \pi, \\[.2cm]
{1 \over 2} \Big ( 
{\rm dilog} \Big ( {|k| - 2 \pi \over 2 \pi} \Big ) -
{\rm dilog} \Big (
{2 \pi \over |k|} \Big ) +
\log \Big ( {|k| - 2 \pi \over 2 \pi} \Big ) \log \Big ( {|k| \over 2 \pi}
\Big ) \Big ), & 2 \pi < |k| < 4 \pi, \end{array} \right.
\nonumber \\
g_4(k) & = & \left \{ \begin{array}{l}
- {\pi^2 \over 6} - {\rm dilog} \Big ( {2 \pi \over 2 \pi - |k|} \Big )
+ 2 {\rm dilog} \Big ( {4 \pi - |k| \over 2 \pi - |k|} \Big )
+ 2 \log \Big ( {2 \pi \over 2 \pi - |k|} \Big )
\log \Big ( {4 \pi - |k| \over 2 \pi - |k|} \Big ), \: \: |k| < 2\pi\\[.2cm]
{\rm dilog} \Big ( {|k| \over |k| - 2 \pi} \Big ) -
2 {\rm dilog} \Big ( {2 \pi \over |k| - 2 \pi} \Big )
+ \log \Big ( {2 \pi \over |k| - 2 \pi} \Big )
\log \Big ( {|k| \over |k| - 2 \pi} \Big ), \: \: 
2\pi < |k| < 4 \pi. \end{array} \right. \nonumber \\
\end{eqnarray}

The above formula for $S(k;\beta)$ in the case $|k| < 2 \pi$ (recall here
$\rho = 1$) can be used to expand $\partial S(k;\beta) / \partial \beta$
about $k=0$. For this task we use computer algebra, which gives the
result
\begin{eqnarray}\label{da}
{\partial S(k;\beta) \over \partial \beta} \bigg |_{\beta = 4} & = &
- {|k| \over 16 \pi} + {|k|^3 \over 256 \pi^3} +
{5 k^4 \over 3072 \pi^4} + {3 |k|^5 \over 4096 \pi^5} +
{27 k^6 \over 81920 \pi^6} \nonumber \\
&& + {37 |k|^7 \over 245760 \pi^7} +
{1273 k^8 \over 18350080 \pi^8} +
{887 |k|^9 \over 27525120 \pi^9} + {4423 k^{10} \over 293601280
\pi^{10}}  \nonumber \\
&& + {1949 |k|^{11} \over 275251200 \pi^{11}} + \cdots
\end{eqnarray}
This allows us to deduce a further equation for $\{b_{8,j}
\}_{j=0,\dots,4}$ and $\{\tilde{a}_{9,j}\}_{j=0,\dots,4}$, which in
combination with (\ref{cw}), (\ref{re4}), (\ref{us1}), (\ref{nt1}) and
(\ref{nt2}) implies
\begin{eqnarray}
p_8(x) & = & (x-1)^2 \Big ( 1 - {263 \over 84} x + {1697 \over 315} x^2
- {6337 \over 1008} x^3 + {1697 \over 315} x^4 - {263 \over 84} x^5
+ x^6 \Big )  \label{p8}\\
p_9(x) & = & (x-1) \Big ( 1 - {791 \over 180} x + 
{73603 \over 7560} x^2 - {7355 \over 504} x^3 +
{2231 \over 135} x^4 - {7355 \over 504} x^5 +
{73603 \over 7560} x^6 - {791 \over 180} x^7 + x^8 \Big ) \nonumber
\\ \label{p9}
\end{eqnarray}

\section{Conclusion}
\setcounter{equation}{0}

Collecting together the evaluations (\ref{6.2}), (\ref{u6h}),
(\ref{p3}), (\ref{g6c}), (\ref{re4'}), (\ref{p7}), (\ref{p8})
and (\ref{p9}), and substituting in (\ref{6}) we have that for
$|k| < {\rm min}\,(2\pi \rho, \pi \beta \rho)$
\begin{eqnarray}\label{pr}
\lefteqn{
{\pi \beta \over |k|} S(k;\beta) =} \nonumber \\
&&
1 \nonumber \\&&
+ (x-1)y \nonumber \\&&
+ (x-1)^2y^2 \nonumber \\&&
+ (x-1) (x^2 - {11 \over 6} x + 1) y^3 \nonumber \\&&
+ (x-1)^2(x^2 - {3 \over 2} x + 1) y^4 \nonumber \\&&
+ (x-1) (x^4 - {91 \over 30} x^3 + {62 \over 15} x^2 - {91 \over 30} x + 1)
y^5 \nonumber \\&&
+ (x-1)^2(x^4 - {37 \over 15} x^3 + {13 \over 4} x^2 - {37 \over 15} x + 1)
y^6 \nonumber \\&&
+ (x-1)(x^6 - {1607 \over 420} x^5 + {2011 \over 280} x^4 -
{911 \over 105} x^3 + {2011 \over 280} x^2 - {1607 \over 420} x + 1) y^7
\nonumber \\&&
+ (x-1)^2(x^6 - {263 \over 84} x^5 + {1697 \over 315} x^4 - {6337 \over
1008} x^3 + {1697 \over 315} x^2 - {263 \over 84} x + 1) y^8
\nonumber \\&&
+  (x-1)  ( x^8 - {791 \over 180} x^7 + 
{73603 \over 7560} x^6 - {7355 \over 504} x^5 +
{2231 \over 135} x^4 - {7355 \over 504} x^3 +
{73603 \over 7560} x^2 - {791 \over 180} x + 1 ) y^9 \nonumber \\&&
+ O(y^{10})
\end{eqnarray}
where $x=\beta/2$ and $y=|k|/\pi\beta \rho$. With the 
coefficient of $y^j$ denoted $p_j(x)$ as has been throughout, we recall 
from our workings above that $p_0(x)$, $p_1(x)$, $p_2(x)$ and $p_4(x)$ have
been calculated for general values of $\beta$. In all other cases the
calculation has relied on the assumption that the $p_j(x)$ are indeed
polynomials. On this point we remark that in such cases, excluding
$j=8$ and 9, we have more data points than is necessary to uniquely
specify $p_j(x)$, assuming it is a polynomial, and our extra data points
are consistent with the explicit forms presented in (\ref{pr}).

We remark that the structure exhibited by (\ref{pr}) is familiar from
the study of exactly solvable two-dimensional lattice models
\cite{Gu99}. In this field one encounters two-variable generating
functions $G(x,y)$ say with series expansions of the form
\begin{equation}\label{cny}
G(x,y) = \sum_{n=0}^\infty H_n(x) y^n
\end{equation}
in which $H_n(x)$ is a rational function, and furthermore the denominator
polynomial in $H_n(x)$ only has a small number of (typically no more than
two) distinct zeros. For example, the two-dimensional Ising model
with couplings $J_1$ ($J_2$) between bonds in the horizontal (vertical)
direction and $x:= \exp(-4J_1/k_BT)$, $y:= \exp(-4J_2/k_BT)$ has for its
spontateous magnetisation the celebrated exact expression (see e.g.~\cite{Ba82})
\begin{equation}\label{cny1}
M(x,y) = \Big ( 1 - {16xy \over (1-x)^2 (1-y)^2} \Big )^{1/8}.
\end{equation}
When written in the form (\ref{cny}) one finds
\begin{equation}\label{cny2}
H_n(x) = {2x P_n(x) \over (1-x)^n}
\end{equation}
where $P_n(x)$ is a polynomial of degree $2n-2$ which satisfies the
functional relation
\begin{equation}\label{cny3}
P_n(x) = x^{2n-2} P_n(1/x).
\end{equation}
As emphasized in \cite{Gu99}, the exact solution (\ref{cny1}) can be
uniquely determined by the functional form (\ref{cny2}), together with
the functional (inversion) relation (\ref{cny3}) and the symmetry
relation $M(x,y) = M(y,x)$. For the structure function of the log-gas
we have no analogue of the symmetry relation and so cannot
characterize (\ref{d2}) this way.

One immediate feature of the polynomials $p_j(x)$ in (\ref{pr}) is that
for $j$ even the polynomial $p_j(-x)$ has all coefficients positive, while
for $j$ odd the polynomial $p_j(-x)$ has all coefficients negative.
Another general feature of the $p_j(x)$ in (\ref{pr}), obtained from
numerical computation, is that all the zeros lie on the unit circle in
the complex $x$-plane.
This can be rigorously determined numerically because
the symmetry (\ref{6.1}) implies that if $x_0$ is a zero of
$p_j(x)$, then so is $1/x_0$, which will be the complex conjugate of
$x_0$ if and only if $|x_0|=1$.

The quantum many body interpretation of (\ref{1}) allows us to give
a physical interpretation to the functional relation
(\ref{5}). As the functional relation
is derived from the integral representation
(\ref{d3}), it is appropriate to recall \cite{Ha95} the physical
interpretation of that formula. In (\ref{d3}), with $\beta/2 = p/q$,
there are $q$ integrals over $x_i \in (0,\infty)$ and $p$
integrals over $y_j \in (0,1)$. The variables $x_i$ can be interpreted as
being rapidities of quasi-particle excitations, while the $y_j$ are
rapidities of quasi-hole excitations. Thus the transformation
$\beta \mapsto 4/\beta$ is equivalent to interchanging $p$ and $q$
and thus the quasi-holes and quasi-particles. In (\ref{d3}) this does not
lead to an integral of the same functional form as before;
although the functional form of the integrand is conserved, apart
 from a renormalization of $k$, the
domain of integration is different for $\{x_i\}$ and $\{y_j\}$. But
with $k$ restricted as in (\ref{d2a}) both sets of variables can take
any value in $(0,\infty)$. The quasi-particles and quasi-holes play an
identical role and the functional equation results.

It is of interest to consider the small $k$ expansion of
$S(k;\Gamma)$, $\Gamma := q^2/k_BT$ ($q$ = charge), for the two-dimensional
one-component plasma. As mentioned earlier, this has recently been the
object of study of Kalinay et al.~\cite{KMST99}. They obtain results which
imply
\begin{equation}\label{sos}
{2 \pi \Gamma  \over k^2} S(k;\Gamma) = 1 +
({\Gamma \over 4}-1)  {k^2 \over 2 \pi \Gamma \rho} +
({\Gamma \over 4} - {3 \over 2})( {\Gamma \over 4} - {2 \over 3})
\Big ( {k^2 \over 2 \pi \Gamma \rho} \Big )^2 + O(k^6).
\end{equation}
where $k := |\vec{k}|$. The structure of (\ref{sos}) bears a striking
resemblence to (\ref{pr}) with $\Gamma/4$ corresponding to $x$ and
$k^2/2\pi \Gamma \rho$ to $y$. In particular with $g(x,y) :=
(2\pi \Gamma / k^2) S(k;\Gamma)$, the expansion (\ref{sos}) to the
given order is such that
\begin{equation}\label{sos1}
g(x,y) = g( {1 \over x}; - yx).
\end{equation}
Furthermore, writing
\begin{equation}\label{sos2}
g(x,y) = 1 + \sum_{l=1}^\infty u_{l}(x) y^{l}
\end{equation}
we have $u_1(x) = (x-1)$, $u_2(x) = (x-3/2)(x-2/3)$ so $u_j(x)$ is a
monic $j$th degree polynomial for $j \le 2$. However we can demonstrate that
this analogy breaks down for the $l=3$ term in (\ref{sos2}).

To demonstrate this fact, suppose instead that the functional equation
(\ref{sos2}) was valid at order $l=3$ in (\ref{sos2}) and $u_3(x)$ is
a monic polynomial. Then $u_3$ must be of the form
\begin{equation}\label{sos3}
u_3(x) = (x-1)(x^2 + ax + 1).
\end{equation}
From the definition of $g(x,y)$ we can check that this is equivalent to
the statement that
\begin{equation}\label{sos4}
{1 \over \rho} \Big ( {\pi \Gamma \rho \over 2} \Big )^4
\int_{{\bf R}^2} {r}^8 S({r};\Gamma) \, d\vec{r} =
(4!)^2(x-1)(x^2 + ax + 1).
\end{equation}
But as noted in \cite{KMST99}, it follows from the perturbation expansion
of \cite{Ja81} that
\begin{eqnarray}\label{sos5}
{1 \over \rho} \Big ( {\pi \Gamma \rho \over 2} \Big )^4
\int_{{\bf R}^2} {r}^8 S({r};\Gamma) \, d\vec{r} & = &
- 4! + (\Gamma - 2)4! \Big ( \sum_{k=0}^4
{2^k - 1 \over k+1} - 2 \Big ) + O((\Gamma - 2)^2) \nonumber \\
& = & - 4! + (\Gamma - 2)4! {17 \over 4} + O((\Gamma - 2)^2).
\end{eqnarray}
The term in (\ref{sos5}) proportional to $\Gamma - 2$ is incompatible
with (\ref{sos4}) which gives instead
$$
(\Gamma - 2)4! {18 \over 4}
$$
independent of the value of $a$.
Indeed in \cite{KMST99} evidence is presented which indicates
$u_3(x)$ is an infinite series in $x$, although we have no way of
determining if the functional equation (\ref{sos1}) also breaks down at
this order.

\section*{Acknowledgements}
The work of PJF and DSM was supported by the Australian Research Council.

\setcounter{equation}{0}
\setcounter{section}{1}
\renewcommand{\thesection}{\Alph{section}}
\section*{Appendix A}
In this appendix some details of the derivation of (\ref{af0}) and (\ref{af1})
will be given. To simplify notation we take $\rho =1$ throughout. The first
step is to substitute (\ref{ny1}) and (\ref{ny0}) and simplify by
expanding out the determinant and cancelling terms where possible. This
shows that up to terms $ O((\beta - 2)^2)$
\begin{eqnarray}\label{fi1}
\lefteqn{
g_2(x_1,x_2;\beta)} \nonumber \\ && = 1 - \Big ( P_2(x_1,x_2) \Big )^2 +
(\beta - 2) \bigg \{ - (1 - (P_2(x_1,x_2))^2) \Phi(x_1,x_2) \nonumber \\&&
- 2 \int_{-\infty}^\infty \Big ( - (P_2(x_2,x_3))^2 -
(P_2(x_1,x_3))^2 + 2 P_2(x_1,x_2) P_2(x_2,x_3) P_2(x_3,x_1) \Big )
\Phi (x_1,x_3) \, dx_3 \nonumber \\&&
- {1 \over 2} \int_{-\infty}^\infty \Big (
4 P_2(x_1,x_3) P_2(x_3,x_4)P_2(x_4,x_1) -
4 P_2(x_1,x_2) P_2(x_2,x_3) P_2(x_3,x_4) P_2(x_4,x_1) \nonumber \\&&
- 2P_2(x_1,x_3) P_2(x_3,x_2) P_2(x_2,x_4) P_2(x_4,x_1) +
2 \Big ( P_2(x_1,x_3) \Big )^2 \Big (P_2(x_2,x_4) \Big )^2 \Big )
\Phi(x_3,x_4) \, dx_3 dx_4 \bigg \} . \nonumber \\
\end{eqnarray}
The convolution structure
$$
\int_{-\infty}^\infty f(y_1 - x) g(x - y_2) \, dx
$$
often occurs in the above integrals. Such an integral can be transformed
by introducing the Fourier transforms $\hat{f}$ ($\hat{g}$) according
to the formula
\begin{equation}\label{fi1'}
\int_{-\infty}^\infty f(y_1 - x) g(x - y_2) \, dx =
{1 \over 2 \pi} \int_{-\infty}^\infty \hat{f}(l) \hat{g}(l)
e^{-il(y_1 - y_2)} \, dl.
\end{equation}
Making use of this formula typically leads to simplifications.

For example, consider the first integral in (\ref{fi1}). Starting with
the Fourier transform
\begin{equation}\label{fi2}
\int_{-\infty}^\infty {\sin^2 \pi x \over (\pi x)^2} e^{ikx} \, dx
= \left \{ \begin{array}{ll} 1 - {|k| \over 2 \pi}, & |k| < 2 \pi \nonumber 
\\[.1cm]
0, & |k| > 2 \pi \end{array} \right.
\end{equation}
and (\ref{fte}), application of (\ref{fi1'}) gives
\begin{equation}\label{fi3}
A_1(x_{12}) := \int_{-\infty}^\infty \Big ( P_2(x_2,x_3) \Big )^2
\Phi (x_3,x_1) \, dx_3 =
\int_{-2 \pi}^{2 \pi} \Big ( 1 - {|k| \over 2 \pi} \Big )
{\pi \over |k|} \cos k x_{12} \, {dk \over 2 \pi}.
\end{equation}
This expression is indeed simpler than the original, but it suffers from
being ill-defined, due to the singularity at the origin. However its
derivative is well-defined, and can furthermore be evaluated in terms
of elementary functions giving
\begin{equation}\label{fi4}
{d \over dx} A_1(x) = {\sin 2 \pi x \over 2 \pi x^2} - {1 \over x}.
\end{equation}
Also, we have \cite{GR80}
\begin{equation}\label{fi4'}
A_1(0) = - \int_{-\infty}^\infty dx \,
{\sin^2 \pi x \over (\pi x)^2} \log |x| = C + \log 2 \pi - 1,
\end{equation}
where $C$ denotes Euler's constant. Together (\ref{fi4}) and (\ref{fi4'})
imply
\begin{equation}\label{fi5}
A_1(x) = - {\sin 2 \pi x \over 2 \pi x} + {\rm ci} \, (2 \pi x) 
- \log |x|
\end{equation}
where ci$(x)$ denotes the cosine integral (\ref{ci}). 

The other six integrals in (\ref{fi1}) yield to similar techniques. We
find
\begin{eqnarray}\label{fi7}
A_2 & := & \int_{-\infty}^\infty \Big ( P_2(x_1,x_3) \Big )^2
\Phi(x_1,x_3) \, dx_3 \: = \: A_1(0) \nonumber \\
A_3(x_{12}) & := & \int_{-\infty}^\infty P_2(x_2,x_3) P_2(x_3,x_1)
\Phi (x_3,x_1) \, dx_3 \: = \:
\int_{-\pi}^\pi \Big ( C + \log(\pi + k) \Big ) \cos kx_{12} \, {dk \over 2 \pi}
\nonumber \\
& = & {1 \over 2} \Big ( C + \log 2 \pi - \log |x_{12}| + {\rm ci} \, 
(2 \pi x_{12})
\Big ) {\sin \pi x_{12} \over \pi x_{12}}
- {1 \over 2} \Big ( {\rm si} \, (2 \pi x_{12}) + {\pi \over 2} \Big )
{\cos \pi x_{12} \over \pi x_{12}} \nonumber \\
A_4 & := & \int_{-\infty}^\infty P_2(x_1,x_3) P_2(x_3,x_4) P_2(x_4,x_1)
\Phi (x_3,x_4) \, dx_3 dx_4 \: = \: A_3(0) \: = \: A_1(0), \nonumber \\
A_5(x_{12}) & := & 
\int_{-\infty}^\infty P_2(x_2,x_3) P_2(x_3,x_4) P_2(x_4,x_1)
\Phi (x_3,x_4) \, dx_3 dx_4 \: = \: A_3(x_{12}), \nonumber \\
A_6(x_{12}) & := & 
\int_{-\infty}^\infty P_2(x_1,x_3) P_2(x_3,x_2) P_2(x_2,x_4)
P_2(x_4,x_1) \Phi(x_3,x_4)\, dx_3 dx_4 \nonumber \\
& = & {1 \over 2 (\pi x_{12})^2} \Big ( C + \log 2 \pi +
\cos 2 \pi x_{12} \Big ( \log |x_{12}| - {\rm ci} \, (2 \pi x_{12}) \Big )
- \sin 2 \pi x_{12}
 \Big ( {\rm si} \, (2 \pi x_{12}) + {\pi \over 2} \Big ) \Big ),
\nonumber \\
A_7(x_{12}) & := & \int_{-\infty}^\infty
\Big ( P_2(x_1,x_3) \Big )^2 \Big ( P_2(x_2,x_4) \Big )^2
\Phi (x_3, x_4) \, dx_3 dx_4 \: = \:
\int_{-2 \pi}^{2 \pi} {dk \over 2 \pi} \,
\Big ( 1 - {|k| \over 2 \pi} \Big )^2 {\pi \over |k|} \cos k x_{12}
\nonumber \\
& = & - \log |x_{12}| - {1 - \cos 2 \pi x_{12} \over (2 \pi x_{12})^2}
- {\sin 2 \pi x_{12} \over 2 \pi x_{12}} + {\rm ci} \, (2 \pi x_{12}),
\end{eqnarray}
where si$(x)$ denotes the sine integral defined in (\ref{si}).

Of the results (\ref{fi7}), the evaluation of $A_6$ is the most difficult,
so it is appropriate to give details in that case also. We observe that
$A_6$ consists of the convolution of $P_2(x_1,x_3) P_2(x_3,x_2)$
regarded as a function of $x_3$, and $\Phi(x_3,x_4)$, and
$P_2(x_4,x_1) P_2(x_2,x_4)$ regarded as a function of $x_4$. It
simplifies the calculation to take as the origin in both integrations
the centre of the interval between particle 1 and particle 2, which is
achieved by the change of variables $x_3 \mapsto x_3 + (x_1 + x_2)/2$,
$x_4 \mapsto x_4 + (x_1 + x_2)/2$. Use of (\ref{fi1'}) then shows
\begin{equation}\label{se1}
A_6(x_{12}) = \int_{-\infty}^\infty {dk \over 2 \pi} \,
\Big ( \hat{V}(k,x_{12}) \Big )^2 {\pi \over |k|},
\end{equation}
\begin{equation}\label{se2}
 \hat{V}(k,x_{12}) := \int_{-\infty}^\infty {\sin \pi x_{13} \over \pi x_{13}}
{\sin \pi x_{32} \over \pi x_{32}} \cos k x_3 \, dx_3 =
\left \{ \begin{array}{ll} {1 \over \pi x_{12}} \sin 
\Big ( \pi - {|k| \over 2} \Big ) x_{12}, & |k| < 2 \pi \\[.2cm]
0, & |k| > 2 \pi \end{array} \right.
\end{equation}
where the second equality in (\ref{se2}) follows after further use of
(\ref{fi1'}). Thus
\begin{equation}\label{se3}
A_6(x) = \int_{-2\pi}^{2\pi} {dk \over 2 \pi} \,
\Big ( {\sin ( \pi - |k| / 2)x \over \pi x}
\Big )^2 {\pi \over |k|}.
\end{equation}
As in (\ref{se3}), this integrand is ill-defined. To proceed further, we write
$$
A_6(x) = A_6^{(1)}(x) + A_6^{(2)}(x)
$$
where
\begin{eqnarray*}
A_6^{(1)}(x) & = & \int_{-\infty}^\infty {dk \over 2 \pi} \,
\bigg \{ \Big ( {\sin ( \pi - |k| / 2)x \over \pi x}
\Big )^2 - \Big ( {\sin \pi x \over \pi x} \Big )^2 \bigg \}
{\pi \over |k|} \nonumber \\
A_6^{(2)}(x) & = &
\Big ( {\sin \pi x \over \pi x} \Big )^2 \int_{-2 \pi}^{2 \pi}
{dk \over 2 \pi} \, {\pi \over |k|}.
\end{eqnarray*}
The integral defining $A_6^{(1)}$ is well defined and can be computed
by elementary means. The integral defining $A_6^{(2)}$ is singular. It
coincides with the singular part of $A_1(0)$ (recall (\ref{fi3})), and
so from (\ref{fi4'}) we have
$$
A_6^{(1)}(0) = \Big ( {\sin \pi x \over \pi x} \Big )^2 
( C + \log 2 \pi ).
$$

Collecting together the above evaluations of $A_1$--$A_7$ and substituting
as appropriate in (\ref{fi1}) gives (\ref{af0}).

 The next task is to evaluate
the Fourier transform. Now the evaluations of $A_1$ and $A_7$ are given
as Fourier integrals, so their Fourier transform is immediate:
\begin{eqnarray}\label{a1a7}
{\rm FT} \, A_1(x) & = & {\pi \over |k|} - {1 \over 2}, \quad |k| < 2\pi
 \nonumber \\
{\rm FT} \, A_7(x) & = & {\pi \over |k|} - 1 + {|k| \over 4 \pi}, 
\quad |k| < 2\pi,
\end{eqnarray}
while for $|k| > 2 \pi$
\begin{equation}\label{a1a7'}
{\rm FT} \, A_1(x) = {\rm FT} \, A_7(x) = 0.
\end{equation}
We can check that the constants
$A_2$ and $A_4$ cancel when substituted in (\ref{fi1}), and so play
no further part in the calculation.

Of the remaining terms, consider first the first term proportional
to $\beta - 2$ in (\ref{fi1}), $A_0(x)$ say. Making use of (\ref{fi1'}) we
see that
\begin{equation}\label{ii}
{\rm FT} \, A_0(x) = - {\pi \over |k|} + 
\int_{-2\pi}^{2\pi} {dl \over 2 \pi} \,
{\pi \over |l-k|} \Big ( 1 - {|l| \over 2 \pi} \Big ).
\end{equation}
For $|k| < 2 \pi$ minor manipulation allows the singular part
\begin{equation}\label{a0a}
\int_{-2\pi}^{2 \pi} {dl \over 2 \pi} \, {\pi \over |l|} = C +
\log 2 \pi
\end{equation}
to be separated, while the remaining convergent integrals are elementary.
We thus find that for $|k| < 2 \pi$
\begin{equation}\label{iia}
{\rm FT} \, A_0(x) = - {\pi \over |k|} + \Big \{ C + \log 2 \pi
+ {1 \over 2} \log \Big ( 1 - \Big ( {k \over 2 \pi} \Big )^2 \Big )
\Big \} \Big ( 1 - {|k| \over 2 \pi} \Big ) -1 + {|k| \over 2 \pi}
+ {|k| \over 2 \pi} \log \Big ( {2 \pi + |k| \over |k|} \Big ).
\end{equation}
For $|k| > 2 \pi$ the integrals in (\ref{ii}) are convergent and also
elementary. In this case we find
\begin{equation}\label{a0b}
{\rm FT} \, A_0(x) = - {\pi \over |k|} + 
{1 \over 2} \log {|k| +2 \pi \over |k| - 2 \pi} +
{|k| \over 4 \pi} \log \Big ( 1 - {4 \pi^2 \over k^2} \Big ).
\end{equation}

To compute the Fourier transform of $A_6$, we begin by making use of
(\ref{fi1'}) in (\ref{se1}) thereby obtaining
$$
{\rm FT} \, A_6 = \int_{-\infty}^\infty {dl \over 2 \pi}
\int_{-\infty}^\infty {dk_1 \over 2 \pi} \, \hat{V}(l,k_1)
\hat{V}(l,-(k_1-k)) {\pi \over |l|}
$$
where
$$
\hat{V}(l,k) := \int_{-\infty}^\infty dx \, \hat{V}(l,x) e^{ikx} =
\chi_{|k| < \pi - |l|/2}
$$
with the equality in the latter formula following from the explicit form
(\ref{se2}) of $\hat{V}(l,x)$ and then computation of the resulting integral,
and where $\chi_T = 1$ for $T$ true and $\chi_T = 0$ otherwise. Thus
\begin{eqnarray}\label{a6}
{\rm FT} \, A_6(x) & = & \int_{-\infty}^\infty {dl \over 2 \pi}
\int_{-\infty}^\infty {dk_1 \over 2 \pi} \,
\chi_{|k_1| < \pi - |l|/2} \chi_{|k_1 - k| < \pi - |l|/2} 
{\pi \over |l|} \nonumber \\
& = & \left \{ \begin{array}{ll}
\Big ( 1 - {|k| \over 2 \pi} \Big ) ( C + \log 2 \pi ) -
 \Big ( 1 - {|k| \over 2 \pi} \Big ) \log {2 \pi \over 2 \pi - |k|}
- {1 \over 2 \pi} (2 \pi - |k|), & |k| < 2 \pi \\
0, & |k| > 2 \pi \end{array} \right. \nonumber \\
\end{eqnarray}
where use has been made of the generalized integral evaluation (\ref{a0a}).

The final Fourier transform to consider is
\begin{eqnarray*}\lefteqn{{\rm FT} \, {\sin \pi x \over \pi x} A_3(x) =
{\rm FT} \, {\sin \pi x \over \pi x}
\int_{-\pi}^\pi {dk_1 \over 2 \pi} \,
\Big ( C + {1 \over 2} \log (\pi + k_1) + {1 \over 2} \log (\pi - k_1)
\Big ) e^{ik_1x}} \\ &&
= {1 \over 2 \pi} \int_{-\infty}^\infty dl \,
\chi_{l \in [-\pi,\pi]} \chi_{l \in [-\pi+k,\pi+k]}
\Big ( C + {1 \over 2} \log (\pi + l-k) + {1 \over 2} \log (\pi - (l-k))
\Big )
\end{eqnarray*}
where to obtain the equality use has been made of (\ref{fi1'}).
Evaluating the integral gives
\begin{equation}\label{a3}
{\rm FT} \, {\sin \pi x \over \pi x} A_3(x) = 
(C + \log 2 \pi) \Big ( 1 - {|k| \over 2 \pi} \Big ) - {|k| \over 4 \pi}
\log {|k| \over 2 \pi} + {1 \over 2} \Big ( 1 - {|k| \over 2 \pi} \Big )
\log \Big ( 1 - {|k| \over 2 \pi} \Big )
\end{equation}
for $|k| < 2 \pi$, while for $|k| > 2 \pi$ 
\begin{equation}\label{a3'}
{\rm FT} \, {\sin \pi x \over \pi x} A_3(x) = 0.
\end{equation}

Substituting the above results as appropriate in the Fourier transform of
(\ref{fi1}) gives the result (\ref{af1}).

\setcounter{equation}{0}
\setcounter{section}{2}
\renewcommand{\thesection}{\Alph{section}}
\section*{Appendix B}
In this appendix we outline some details of the calculation of (\ref{ny0})
in the case $\beta_0=4$ and show how this leads to (\ref{b4}). Because
(\ref{ny1}) and (\ref{ny2}) formally have the same structure upon expansion
(recall the definition of qdet (\ref{qdet})), the formula (\ref{fi1})
formally maintains its structure when generalized to the case $\beta_0=4$.
Thus we have
\begin{eqnarray}\label{gi1}
\lefteqn{
g_2(x_1,x_2;\beta)} \nonumber \\ && = 
1 - \Big ( P_4(x_1,x_2) P_4(x_2,x_1) \Big )^{(0)} +
(\beta - 4) \bigg \{ - \Big (1 - 
(P_4(x_1,x_2) P_4(x_2,x_1))^{(0)} \Big ) \Phi(x_1,x_2) \nonumber \\&&
- 2 \int_{-\infty}^\infty \Big ( - (P_4(x_2,x_3) P_4(x_3,x_2))^{(0)} -
(P_4(x_1,x_3) P_4(x_3,x_1))^{(0)} \nonumber \\&& + 
2 (P_4(x_1,x_2) P_4(x_2,x_3) P_4(x_3,x_1))^{(0)} \Big )
\Phi (x_1,x_3) \, dx_3 \nonumber \\&&
- {1 \over 2} \int_{-\infty}^\infty \Big (
4 (P_4(x_1,x_3) P_4(x_3,x_4)P_4(x_4,x_1))^{(0)} -
4 (P_4(x_1,x_2) P_4(x_2,x_3) P_4(x_3,x_4) P_4(x_4,x_1))^{(0)} \nonumber \\&&
- 2(P_4(x_1,x_3) P_4(x_3,x_2) P_4(x_2,x_4) P_4(x_4,x_1))^{(0)}  \nonumber \\&&
+ 2( P_4(x_1,x_3) P_4(x_3,x_1))^{(0)} (P_4(x_2,x_4)P_4(x_4,x_2))^{(0)} \Big )
\Phi(x_3,x_4) \, dx_3 dx_4 \bigg \} +O((\beta - 4)^2). 
\end{eqnarray}
We treat each of the seven distinct integrals in (\ref{gi1}) in an
analogous way to their counterparts in (\ref{fi1}), although extra working
is involved due to $P_4$ being a matrix rather than a scalar.

The final results are
\begin{eqnarray}\label{B}
B_1(x_{12}) & := & \int_{-\infty}^\infty
\Big ( P_4(x_2,x_3) P_4(x_3,x_2) \Big )^{(0)} \Phi(x_1,x_3) \, dx_3 
\nonumber \\
& = & \int_{-4\pi}^{4\pi} {dk \over 2 \pi} \,
\Big ( 1 - {|k| \over 4 \pi} + {|k| \over 8 \pi}
\log \Big | 1 - {|k| \over 2 \pi} \Big | \Big ) {\pi \over |k|}
\cos k x_{12} \nonumber \\
& = & - \log |x_{12}| - {\sin 4 \pi x_{12} \over 4 \pi x_{12}}
+{\rm ci} \, (4 \pi x_{12}) - {\cos 2 \pi x_{12} \over 4 \pi x_{12}}
{\rm Si} \, (2 \pi x_{12}) \nonumber \\
B_2 & := & \int_{-\infty}^\infty
\Big ( P_4(x_1,x_3) P_4(x_3,x_1) \Big )^{(0)} \Phi(x_1,x_3) \, dx_3 
\: = \: B_1(0) \: = \: C + \log 4 \pi - {3 \over 2}  \nonumber \\
B_3(x_{12}) & := & \int_{-\infty}^\infty
P_4(x_2,x_3) P_4(x_3,x_1)  \Phi(x_1,x_3) \, dx_3 
\nonumber \\
& = &
\left [ \begin{array}{cc}
{1 \over 4} f_1(x_{12}) - {1 \over 4} f_3(x_{12}) &
- {1 \over 4}  \int_0^{x_{12}} (f_1(t) + f_2(t)) \, dt \nonumber \\
- {1 \over 4}f_1'(x_{12}) + {1 \over 4} f_3'(x_{12}) &
{1 \over 4} f_1(x_{12}) + {1 \over 4} f_2(x_{12}) \end{array} \right ]
\nonumber \\
B_4  & := & \int_{-\infty}^\infty 
\Big ( P_4(x_1,x_3) P_4(x_3,x_4) 
P_4(x_4,x_1) \Big )^{(0)} \Phi(x_1,x_3) \, dx_3 dx_4 
\: = \: B_1(0) \nonumber \\
B_5(x_{12}) & := & \int_{-\infty}^\infty
P_4(x_2,x_3) P_4(x_3,x_4) P_4(x_4,x_1) \Phi(x_3,x_4) \, dx_3 dx_4
\nonumber \\
& = &
\left [ \begin{array}{cc}
{1 \over 4} f_1(x_{12}) + {1 \over 8} f_2(x_{12}) 
- {1 \over 8} f_3(x_{12}) &
-   \int_0^{x_{12}} ({1 \over 4}f_1(t) + {1 \over 8} f_2(t)-
{1 \over 8} f_3(t)) \, dt \nonumber \\
- ({1 \over 4}f_1'(x_{12}) + {1 \over 8} f_2'(x_{12})
- {1 \over 8} f_3'(x_{12}))
 &
{1 \over 4} f_1(x_{12}) + {1 \over 8} f_2(x_{12})
- {1 \over 8} f_3(x_{12}) \end{array} \right ]
\nonumber \\
B_6(x_{12}) & := & \int_{-\infty}^\infty \Big (
P_4(x_1,x_3) P_4(x_3,x_2) P_4(x_2,x_4)P_4(x_4,x_3)
\Big )^{(0)} \Phi(x_3,x_4) \, dx_3 dx_4
\nonumber \\
&=& \int_{-4\pi}^{4\pi} {1 \over 2 |k|} \bigg \{ (g_1(k,x_{12}))^2
+ \cos(kx_{12}/2)g_1(k,x_{12})\Big (\frac{|k|}{4\pi}g_2(k,x_{12})
-\frac{ik}{4\pi}g_3(k,x_{12})\Big ) \nonumber \\
&& + \cos(kx_{12})\left(\frac{|k|}{8\pi}g_2(k,x_{12})
-\frac{ik}{8\pi}g_3(k,x_{12})\right)^2
-\Big (\frac{\sin(|k|x_{12}/2)}{4\pi}g_2(k,x_{12})
\nonumber \\&& +\frac{\cos(kx_{12}/2)}{4\pi}g_3(k,x_{12})\Big ) 
\Big (\frac{(4\pi-|k|)\cos((2\pi -|k|/2))x_{12})}{4\pi x_{12}}
\nonumber \\&&
-\frac{\sin((2\pi-|k|/2))x_{12})}
{2\pi x_{12}^2}\Big ) \bigg \} \,dk \nonumber \\
B_7(x_{12}) & := & \int_{-\infty}^\infty \Big (
P_4(x_1,x_3) P_4(x_3,x_1) \Big )^{(0)} \Big ( P_4(x_2,x_4)P_4(x_4,x_2)
\Big )^{(0)} \Phi(x_3,x_4) \, dx_3 dx_4
\nonumber \\
&=&  
\int_{-4\pi}^{4\pi} \left(1-\frac{|k|}{4\pi}
+\frac{|k|}{8\pi}\ln\left|1-\frac{|k|}{2\pi}\right|\right)^2 \frac{\pi}{|k|}
\cos kx_{12} \frac{dk}{2\pi} \nonumber \\
& = & -\log |x| - \frac{\sin 4\pi x}{4\pi x} + {\rm ci} \,(4\pi x)
- \frac{{\rm Si} \, (2\pi x) \sin 2\pi x}{8\pi^2x^2}
-\frac{{\rm Si} (2\pi x) \cos 2\pi x}{4\pi x} \nonumber \\
&& + \int_{-4\pi}^{4\pi} |k| \log \Big |1-\frac{|k|}{2\pi}\Big |^2 \cos kx
\frac{dk}{128\pi^2}
\end{eqnarray}
where
\begin{eqnarray}
f_1(x) & := & 
\int_{-2\pi}^{2\pi} (2C+\ln(4\pi^2-k^2)) \cos kx \frac{dk}{2\pi} \nonumber \\
&=& \frac{\sin 2\pi x}{\pi x}(C+\ln 4\pi-\ln|x|+{\rm ci}(4\pi x))
- \frac{\cos 2\pi x}{\pi x} {\rm Si}(4\pi x) \nonumber \\
f_2(x) & := & - {1 \over \pi x} {\rm Si}\,(2 \pi x), \qquad
f_3(x) \: := \: {\sin 2 \pi x \over \pi x} \nonumber \\
g_1(k,x) & = & {\sin((2\pi - |k|/2)x) \over 2 \pi x} \nonumber \\
g_2(k,x) & = & {\rm ci}\,((2\pi - |k|)x) - {\rm ci}\,(2\pi x) \nonumber \\
g_3(k,x) & = & {\rm Si}\,((2\pi - |k|)x) + {\rm Si}\,(2\pi x) 
\end{eqnarray}

When substituted in (\ref{gi1}), the constant terms $B_2$ and $B_4$ cancel,
and we obtain the formula
\begin{eqnarray}\label{Bg}
g(x_1,x_2;\beta) & = & 1 - \Big (P_4(x_1,x_2)P_4(x_2,x_1) \Big )^{(0)}
\nonumber \\
&& + (\beta - 4) \bigg \{ \Big ( -1 + (P_4(x_1,x_2)P_4(x_2,x_1))^{(0)}
 \Phi(x_1,x_2) \nonumber \\
&& +2 B_1(x_{12}) - 4 \Big ( P_4(x_{12}) B_3(x_{12}) \Big )^{(0)} +
2  \Big ( P_4(x_{12}) B_5(x_{12}) \Big )^{(0)} \nonumber \\
&& + B_6(x_{12}) - B_7(x_{12})  \bigg \} +
O((\beta - 4)^2).
\end{eqnarray}

We will demonstrate the close analogy with the $\beta =2$ calculation of
Appendix A by giving the derivation of the integral formula in
(\ref{B}) for $B_1(x_{12})$. As in the derivation of the integral
formula (\ref{fi3}) for $A_1(x_{12})$, our strategy is to use the
convolution formula (\ref{fi1'}). However here the Fourier transform of
$P_4(x_1,x_2) P_4(x_2,x_1)$ is not immediate. What is immediate is the
Fourier transform of $P_4(x_1,x_2)$. Thus from the definition
(\ref{ny2'}) we see that
\begin{eqnarray*}
{\rm FT} \, P_4(x_1,x_2) & := & \int_{-\infty}^\infty P_4(x_1,x_2)
e^{ikx_{12}} \, dx_{12} \nonumber \\
& = & \left \{ \begin{array}{ll} \left [ \begin{array}{cc} 1/2 & i/2k \\
-ik/2 & 1/2 \end{array} \right ], & |k| < 2 \pi \\
\left [ \begin{array}{cc} 0 & 0 \\ 0 & 0 
\end{array} \right ], & |k| > 2 \pi. \end{array} \right.
\end{eqnarray*}
Use of (\ref{fi1'}) then shows that for $|k| < 4 \pi$
\begin{eqnarray}\label{fi2a}
{\rm FT} \, P_4(x_1,x_2)P_4(x_2,x_1) & := & 
\int_{-2\pi}^{2\pi}
\left [ \begin{array}{cc} 1/2 & i/2l \\
-il/2 & 1/2 \end{array} \right ]
\left [ \begin{array}{cc} 1/2 & -i/2(k-l) \\
i(k-l)/2 & 1/2 \end{array} \right ] \chi_{|k-l|<2\pi} \,
{dl \over 2 \pi} \nonumber \\
& = &
\left [ \begin{array}{cc}
1 - {|k| \over 4 \pi} + {|k| \over 8 \pi} \log \Big |
1 - {|k| \over 2 \pi} \Big | & 0  \\
0 & 1 - {|k| \over 4 \pi} + {|k| \over 8 \pi} \log \Big |
1 - {|k| \over 2 \pi} \Big | \end{array} \right ],
\end{eqnarray}
while for $|k| > 4 \pi$
\begin{equation}\label{fi2b}
{\rm FT} \, P_4(x_1,x_2) P_4(x_2,x_1) = 0.
\end{equation}
The results (\ref{fi2a}) and (\ref{fi2b}) are the analogue of (\ref{fi2})
in the working leading to the evaluation of $A_1(x_{12})$. The integral
formula for $B_1(x_{12})$ in (\ref{B}) now follows from (\ref{fi2a}),
(\ref{fi2b}) and (\ref{fte}) upon a further application of (\ref{fi1'}).

The Fourier transform of (\ref{Bg}) can be computed explicitly. The final
result has already been stated in (\ref{b4}). This is obtained through the
intermediate results
$$
{\rm FT} \, B_j(x) = \hat{B}_j(k) \quad {\rm for} \quad j=0,1,3,5,6,7
$$
with
$$
B_0(x_{12}) := \Big ( P_4(x_1,x_2) P_4(x_2,x_1) \Big )^{(0)}
\Phi(x_1,x_2)
$$
and the $\hat{B}_j$ specified by (\ref{B0})--(\ref{B7}). We will
illustrate the working by giving some details of the computation
of $\hat{B}_0(k)$ for $|k| < 4 \pi$.

Using (\ref{fi2a}) and (\ref{fte}) we see from (\ref{fi1'}) that
\begin{eqnarray}\label{see}
{\rm FT} \, B_0(x_{12}) & = &
\int_{-4\pi}^{4\pi} \Big (
1 - {|l| \over 4 \pi} + {|l| \over 8 \pi}
\log \Big | 1 - {|l| \over 2 \pi} \Big | \Big ) {\pi \over |k-l|} \,
{dl \over 2 \pi} \nonumber \\
& = & \Big (
1 - {|k| \over 4 \pi} + {|k| \over 8 \pi}
\log \Big | 1 - {|k| \over 2 \pi} \Big | \Big ) \int_{-4\pi}^{4\pi}
{\pi \over |k-l|} \, {dl \over 2 \pi} \nonumber \\
&& + \int_{-4\pi}^{4\pi} \Big ( {|k| - |l| \over 4 \pi} \Big )
{\pi \over |k-l|} \,{dl \over 2 \pi} +
\int_{-4\pi}^{8 \pi} \Big ( {|l| - |k| \over 8 \pi} \Big )
\log \Big | 1 - {|l| \over 2 \pi} \Big |
{\pi \over |k-l|} \,{dl \over 2 \pi} \nonumber \\
&& + {|k| \over 8 \pi}
\int_{-4\pi}^{4\pi} \log \Big | {2 \pi - |l| \over 2 \pi - |k|} \Big |
{\pi \over |k-l|} \,{dl \over 2 \pi}
\end{eqnarray}
where the second equality, which follows from minor manipulation of the
first integral, is motivated by the desire to separate the singular integral.
Thus in the second equality of (\ref{see}) only the first integral is
singular. It is essentially the same as the first singular integral in
(\ref{ii}), and is evaluated as
\begin{equation}\label{nw1}
\int_{-4\pi}^{4\pi}
{\pi \over |k-l|} \, {dl \over 2 \pi} = C + {1 \over 2}
\log (16 \pi^2 - k^2), \qquad |k| \le 4 \pi.
\end{equation}

The second integral in the second equality of (\ref{see}) also appears in
the evaluation of (\ref{ii}). An elementary calculation shows
\begin{equation}\label{nw2}
\int_{-4\pi}^{4\pi} {|k| - |l| \over 4 \pi} {\pi \over |k-l|} \,
{dl \over 2 \pi} = - 1 + {|k| \over 4 \pi} + {|k| \over 4 \pi}
\log \Big ( {4\pi + |k| \over |k|} \Big ).
\end{equation}

To evaluate the third integral in (\ref{see}) we suppose without loss of 
generality that $k>0$ and write
\begin{eqnarray}\label{nw3}
\lefteqn{\int_{-4\pi}^{4\pi} {|l| - |k| \over 8 \pi}
\log \Big | 1 - {|l| \over 2 \pi} \Big | {\pi \over |k-l|} \,
{dl \over 2 \pi}  = \int_0^{2\pi} \Big ( {1 \over 16 \pi} + {k \over 8 \pi (l-k)} \Big )
\log \Big | 1 + {l \over 2 \pi} \Big | \, dl
}  \nonumber \\  &&
 - \int_0^k \log \Big | 1 - {|l| \over 2 \pi} \Big | \, {dl \over 16 \pi}
+ \int_k^{4 \pi}  \log \Big | 1 - {|l| \over 2 \pi} \Big | \, {dl \over 16 \pi}.
\end{eqnarray}
The only non-elementary integral is the second term of the first integral.
This can be computed by checking from the definition (\ref{dilog})
that for $-4\pi < l < 0$
\begin{equation}\label{nw3'}
{d \over dl} \bigg (
{\rm dilog} \, \Big ( {k-l \over k + 2\pi} \Big ) +
\log \Big | 1 + {l \over 2 \pi} \Big |
{\rm log} \, \Big ( {k-l \over k + 2\pi} \Big ) \bigg ) =
{1 \over l - k} \log \Big | 1 + {l \over 2 \pi} \Big |.
\end{equation}
In total we therefore have
\begin{eqnarray}\label{nw4}
\lefteqn{
\int_{-4\pi}^{4\pi} {|l| - |k| \over 8 \pi}
\log \Big | 1 - {|l| \over 2 \pi} \Big | {\pi \over |k-l|} \,
{dl \over 2 \pi}} \nonumber \\
&& = - {1 \over 2} + {|k| \over 8 \pi} +
{|k| \over 8 \pi} \Big ( {\rm dilog} \, \Big ( {|k| \over 2 \pi + |k|} \Big )
- {\rm dilog} \, \Big ( {4 \pi + |k| \over 2 \pi + |k|} \Big ) \Big )
+ {2 \pi - |k| \over 8 \pi} \log \Big | 1 - {|k| \over 2 \pi} \Big |.
\end{eqnarray}

To evaluate the final integral in (\ref{see}), a similar approach to that
leading to the evaluation (\ref{nw4}) is adopted. Minor complications
arise because of the need to modify the formula (\ref{nw3'}) for
$l > k$. We find
\begin{eqnarray}\label{nw5}
\lefteqn{
\int_{-4\pi}^{4\pi} {|k| \over 8 \pi} \log \Big | {2 \pi - |l| \over
2 \pi - |k|} \Big | {\pi \over |k-l|} \, {dl \over 2 \pi}} \nonumber \\
&& = {|k| \over 16 \pi} \bigg \{
{\rm dilog} \, \Big ( {4 \pi + |k| \over 2 \pi + |k|} \Big ) -
{\rm dilog} \, \Big ( {|k| \over 2 \pi + |k|} \Big ) -
\log \Big | 1 - {|k| \over 2 \pi} \Big | 
\log \Big ( {4 \pi + |k| \over |k|} \Big ) + g_1(k) \bigg \}  \nonumber \\
\end{eqnarray}
where $g_1$ is defined in (\ref{gg}). Substituting (\ref{nw1})--(\ref{nw5})
in (\ref{see}) gives the result 
(\ref{B0}).


\end{document}